\documentclass[aps,prl,reprint,superscriptaddress,showpacs,nofootinbib]{revtex4-1}

\usepackage{hyperref}
\usepackage{amsmath}
\usepackage{amssymb}
\usepackage{graphicx}
\usepackage{color}
\usepackage{subfig}
\usepackage{bm}
\usepackage{gensymb}
\usepackage{bigints}
\usepackage[usenames,dvipsnames,svgnames,table]{xcolor}
\usepackage{pifont}
\usepackage{aas_macros}
\usepackage{orcidlink}
\usepackage{physics}

\usepackage[labelfont=bf]{caption}

\hypersetup{
    colorlinks=true,
    linkcolor=Blue,
    filecolor=Blue,
    urlcolor=MidnightBlue,
    citecolor=Blue,
   pdftitle={Cooling age}
}

\begin{document}


\title{Bose-Einstein Condensate and Liquid Helium He$^4$: Implications of GUP and Modified Gravity Correspondence}


\author{Aneta Wojnar\orcidlink{0000-0002-1545-1483}}
\thanks{Corresponding author}
\email[E-mail: ]{awojnar@ucm.es}
\affiliation{Department of Theoretical Physics \& IPARCOS, Complutense University of Madrid, E-28040, 
Madrid, Spain}

\begin{abstract}
Utilizing the recently established connection between Palatini-like gravity and linear Generalized Uncertainty Principle (GUP) models, we have formulated an approach that facilitates the examination of Bose gases. Our primary focus is on the ideal Bose-Einstein condensate and liquid helium, chosen as illustrative examples to underscore the feasibility of tabletop experiments in assessing gravity models. The non-interacting Bose-Einstein condensate imposes constraints on linear GUP and Palatini $f(R)$ gravity (Eddington-inspired Born-Infeld gravity) within the ranges of $-10^{12}\lesssim\sigma\lesssim 3\times 10^{24}{\text{ s}}/{\text{kg m}}$ and $-10^{-1}\lesssim\bar\beta\lesssim 10^{11} \text{ m}^2$ ($-4\times10^{-1}\lesssim\epsilon\lesssim 4\times 10^{11} \text{ m}^2$), respectively.
In contrast, the properties of liquid helium suggest more realistic bounds, specifically $-10^{23}\lesssim\sigma\lesssim 10^{23}{\text{ s}}/{\text{kg m}}$ and $-10^{9}\lesssim\bar\beta\lesssim 10^{9} \text{ m}^2$. Additionally, we argue that the newly developed method employing Earth seismic waves provides improved constraints for quantum and modified gravity by approximately one order of magnitude.
\end{abstract}

\maketitle

\section{Introduction}

Exploring theories beyond General Relativity (GR) becomes imperative despite its success in explaining phenomena, from Solar System dynamics to gravitational waves' detection \cite{abbott2016observation,abbott2017gw170817}. GR faces challenges in accounting for dark matter \cite{rubin1980rotational}, dark energy \cite{huterer1999prospects}, and early cosmological inflation \cite{copeland2006dynamics,nojiri2007introduction}. To address these, Modified Gravity (MG) theories seem to be crucial, offering insights into fundamental cosmic phenomena and uncovering untested regions in the gravitational parameter space \cite{baker2015linking}.

MG introduces alterations impacting various microphysical aspects. Theories suggest changes in chemical potential \cite{kulikov1995low}, geodesic deviation equations on stars' surfaces with a clear microphysical interpretation \cite{kim2014physics}, and microscopic quantities like opacity or specific heat \cite{sakstein2015testing,Kalita:2022trq}. Gravitational proposals affect laws governing thermodynamics, stellar stability, heat transport, and Fermi gas properties \cite{Wojnar:2016bzk,Wojnar:2017tmy,Sarmah:2021ule,Wojnar:2020txr,Guerrero:2021fnz,Gomes:2022sft,wojnar2023fermi}.
Theoretical descriptions of stellar thermonuclear processes, elementary particle interactions, and chemical reaction rates are influenced by gravity modifications \cite{sakstein2015hydrogen,Olmo:2019qsj,crisostomi2019vainshtein,rosyadi2019brown,wojnar2021lithium,lecca2021effects}. Neglecting relativistic effects in equations of state leads to underestimation of compact star limiting masses, with additional changes when (pseudo-)scalar fields are considered \cite{hossain2021equation,hossain2021higher,li2022we,chavanis2004statistical,sakstein2022axion}.

Generalized Uncertainty Principle (GUP) models, involving constants representing the speed of light and gravity, introduce corrections in equations of state and microscopic variables \cite{moussa2015effect,rashidi2016generalized,belfaqih2021white,mathew2021existence,hamil2021new,gregoris2022chadrasekhar}. Integrating the quantum structure of space-time with GUP emphasizes the generalization of the Heisenberg uncertainty principle, offering potential measurable effects \cite{Pachol:2023bkv,Pachol:2023tqa,Kozak:2023vlj}. GUP has proven valuable in predicting quantum gravity effects \cite{kempf1995hilbert,maggiore1993generalized,maggiore1994quantum,chang2002exact,chang2002effect,moussa2015effect,barca2022comparison}, often featuring a minimum length scale around the Planck length, $L_P\sim \sqrt{\frac{\hbar G}{c^3}}$ \cite{bishop2020modified,bishop2022subtle,segreto2023extended}.

The recently established connection between MG and GUP \cite{Wojnar:2023bvv} paves the way for testing gravitational theories in Earth laboratories\footnote{Simultaneously, methods developed by either of these communities can be employed to assess MG or GUP proposals.}.
To illustrate how modified gravity can undergo testing in tabletop experiments, our focus will be on Bose gases. We will develop a general formalism for our deformed phase space and subsequently delve into the specifics of liquid helium, examining its portrayal as a two-fluid model near absolute zero, as presented by Landau. Before that, we will recall the basic notions related to the Ricci-based gravity, its relativistic limit, and the mentioned correspondence resulting as a deformed phase space. The last part of the paper is devoted to our conclusions and future plans.

\section{Deformed phase space in Ricci-based gravity}

In the subsequent discussion, we revisit the connection between modified gravity and the GUP. Initially, we will delve into fundamental concepts associated with Ricci-based gravity. Subsequently, we will revisit the established relationship, exploring the deformation of the phase space in Palatini-like proposals, along with its implications for thermodynamics.

\subsection{Ricci-based gravity}
This specific class of metric-affine theories of gravity is characterized by the following action:

\begin{equation} \label{eq:actionRBG}
\mathcal{S}=\int d^4 x \sqrt{-g} \mathcal{L}_G(g_{\mu\nu},R_{\mu\nu}) +
\mathcal{S}_m(g_{\mu\nu},\psi_m) \ .
\end{equation}

Here, $g$ is the determinant of the space-time metric $g_{\mu \nu}$, and $R_{\mu \nu}$ is the symmetric Ricci tensor, independent of the metric and constructed solely with the affine connection $\Gamma \equiv \Gamma_{\mu\nu}^{\lambda}$. The object ${M^ \mu}_{\nu} \equiv g^{\mu \alpha}R_{\alpha\nu}$ is introduced to formulate the gravitational Lagrangian $\mathcal{L}_G$ as a scalar function using powers of traces of ${M^ \mu}_{\nu}$.

The matter action is given by:

\begin{equation}
\mathcal{S}_m=\int d^4 x \sqrt{-g} \mathcal{L}_m(g_{\mu\nu},\psi_m).
\end{equation}

In this framework, the matter action is minimally coupled to the metric, disregarding the antisymmetric part of the connection (torsion), similar to the treatment of minimally coupled bosonic fields. This simplification extends to fermionic particles, such as degenerate matter, described effectively by a fluid approach exemplified by the perfect fluid energy-momentum tensor \cite{alfonso2017trivial}. By focusing on the symmetric part of the Ricci tensor, potential ghostlike instabilities are avoided \cite{Borowiec:1996kg,Allemandi:2004wn,beltran2019ghosts,jimenez2020instabilities}. This approach accommodates various gravity theories, including GR, Palatini $f(R)$ gravity, Eddington-inspired Born-Infeld (EiBI) gravity \cite{vollick2004palatini}, and its extensions \cite{jimenez2018born}.

The gravitational action encompasses theories that, despite intricate field equations, can be conveniently reformulated, as shown in \cite{jimenez2018born}:

\begin{equation} \label{eq:feRBG}
{G^\mu}_{\nu}(q)=\frac{\kappa}{\vert \hat{\Omega} \vert^{1/2}} \left({T^\mu}_{\nu}-\delta^\mu_\nu \left(\mathcal{L}_G + \frac{T}{2} \right) \right) \ .
\end{equation}
Here, $\vert\hat{\Omega}\vert$ is the determinant of the deformation matrix, and $T$ is the trace of the energy-momentum tensor of matter fields. The Einstein tensor ${G^\mu}_{\nu}(q)$ is associated with a tensor $q_{\mu\nu}$, where the connection $\Gamma$ assumes the Levi-Civita connection of $q_{\mu\nu}$:

\begin{equation}
\nabla_{\mu}^{\Gamma}(\sqrt{-q} q^{\alpha \beta})=0.
\end{equation}

For this formalism, the tensor $q_{\mu\nu}$ is related to the space-time metric $g_{\mu\nu}$ through:

\begin{equation}\label{eq:defmat}
q_{\mu\nu}=g_{\mu\alpha}{\Omega^\alpha}_{\nu} ,
\end{equation}

The deformation matrix ${\Omega^\alpha}_{\nu}$ is theory-dependent, determined by the gravitational Lagrangian $\mathcal{L}_G$. Importantly, these theories yield second-order field equations, reducing to GR counterparts in vacuum (${T_\mu}^{\nu}=0$), implying no extra degrees of freedom propagate in these theories beyond the usual two polarizations of the gravitational field.

{
In what follows, we will focus on two particular theories of modified gravity: Palatini $f(R)$ and EiBI, being at the same time the most studied in the context of the Ricci-based family. Let us compare them on the gravitational action and then field equations levels. Their actions are, respectively
\begin{align} 
S_{Pal}[g,\Gamma,\psi_m] &= \frac{1}{2 \kappa^2} \int d^4x \sqrt{-g} f({R}) + S_m[g_{\mu\nu},\psi_m]\\
{S}_{EiBI}[g,\Gamma,\psi_m]&=\frac{1}{\kappa^2 \epsilon}\int d^4x\left[ \sqrt{-|g_{\mu\nu}+\epsilon {R}_{(\mu\nu)}(\Gamma)|}\right.\nonumber\\
 &\left.-\lambda\sqrt{-g} \right]+
 \mathcal{S}_m[g,\psi_m],\label{eibi}
\end{align}
for which we are interested in the analytic functional in the case of the Palatini $f(R)$ gravity
\begin{equation}
    f(R)=\sum_{i=0}\alpha_i{R}^i.
\end{equation}
Note that expanding the action \eqref{eibi} for fields $ \vert \mathcal{R}_{\mu\nu} \vert \ll \epsilon^{-1}$ yields \cite{pani2012surface},
\begin{align} 
{S}_{EiBI}&=\frac{1}{\kappa^2}\int d^4x\sqrt{g}\left[{R}-2\Lambda+\frac{\epsilon}{4}({R}^2-2{R}_{\mu\nu}{R}^{\mu\nu}) \right.\nonumber\\
&\left.+\mathcal{O}(\epsilon^2)\right]+{S}_m
\end{align}
which essentially describes GR with an effective cosmological constant term $\Lambda=\frac{\lambda-1}{\epsilon}$ and supplemented by quadratic curvature corrections (in the Palatini sense).\\
The field equations are, respectively
\begin{align}
    f'({R}){R}_{\mu\nu}-\frac{1}{2}f({R})g_{\mu\nu}=&\kappa^2 T_{\mu\nu},\\
\nabla_{\mu}^{\Gamma}(\sqrt{-q} q^{\alpha \beta})=&0
\end{align}
for Palatini $f(R)$ gravity with $q_{\mu\nu}=f'({R})g_{\mu\nu}$, and
\begin{align}
     \frac{\sqrt{|q|}}{\sqrt{|g|}}q^{\mu\nu}-\lambda g^{\mu\nu}=&-\epsilon\kappa^2 T^{\mu\nu},\\
\nabla_{\mu}^{\Gamma}(\sqrt{-q} q^{\alpha \beta})=&0 ,
\end{align}
for EIBI one with $q_{\mu\nu}=g_{\mu\nu}+\epsilon\mathcal{R}_{\mu\nu} $. Therefore, the field equations of both theories can be rewritten in the form of \eqref{eq:feRBG} which significantly simplify the computational studies. Among other things, this formalism is used to obtain the non-relativistic limit of those theories. As it can be shown, in Palatini $f(R)$ \cite{Toniato:2019rrd} and EiBI \cite{banados2010eddington,pani2011compact} gravities, the Poisson equation takes the form:
\begin{equation}\label{poisson}
\nabla^2\phi = \frac{\kappa}{2}\Big(\rho+\bar\alpha\nabla^2\rho\Big)
\end{equation}
Here, $\phi$ is the gravitational potential, $\kappa=8\pi G$, and $\bar\alpha$ is a theory parameter. The expressions for $\bar\alpha$ are $\bar\alpha=2\bar\beta$ for Palatini $f(R)$, with $\bar\beta$ accompanying the quadratic term, and $\bar\alpha=\epsilon/2$ for EiBI, where $\epsilon=1/M_{BI}$ and $M_{BI}$ is the Born-Infeld mass. The similarity in the Poisson equation between these two gravity proposals is not coincidental; the EiBI gravity in the first-order approximation reduces to Palatini gravity with the quadratic term \cite{pani2012surface} as recalled above. Furthermore, only the quadratic term $R^2$ influences the non-relativistic equations, as higher curvature scalar terms enter the equations at the sixth order \cite{Toniato:2019rrd}.
}

\subsection{Deformed phase space and resulting thermodynamics}

As demonstrated in \cite{Wojnar:2023bvv}, the additional term appearing in the Poisson equation \eqref{poisson} can be treated as a modification to the Fermi gas for a finite temperature. However, such a modification can be obtained when we deal with a deformation of the phase space
\begin{equation}\label{sumint}
 \frac{1}{(2\pi \hbar)^3} \int \frac{d^3xd^3p}{(1-\sigma p)^{d}},
\end{equation}
in which the subcase $d=1$ refers to the Palatini-like theories of gravity. 
The relation between the deformation parameter $\sigma$ and Palatini parameter $\bar\beta$ is given as follows:
\begin{equation}
\sigma = \frac{4\pi G}{K_2}\bar\beta\,\,\,\,\text{and}\;\;\; K_2 = \frac{3}{\pi} \frac{h^3N_A^2}{m_e \mu_e^2},
\end{equation}
where $m_e$ is the electron mass, $\mu_e$ is the mean molecular weight per electron, and other constants have their usual meaning.

This correspondence allows us to write a general partition function in three dimensions in a large volume as follows:
\begin{equation}\label{partition}
\mathrm{ln}Z = \frac{V}{(2\pi \hbar)^3}\frac{g}{a}\int \mathrm{ln}\left[1-az e^{-E/k_BT}\right] \frac{d^3p}{(1-\sigma p)^{d}} \ ,
\end{equation}
where $V:=\int d^3x$ represents the volume of the cell in configuration space while taking $a=1$ ($a=-1$) one will deal a system of fermionic (bosonic) particles with energy states $E_p$. The fugacity is given by $z=e^{\mu/k_BT}$, the symbol $\mu$ represents the chemical potential while $g$ is a spin of a particle.

In a manner akin to the GUP featuring linear $p$-modifications \cite{cortes2020deformed,ali2009discreteness,ali2011minimal,abac2021modified,vagenas2019gup,tawfik2014generalized}, our methodology incorporates a deformed phase space measure characterized by the deformation parameter $\sigma$. In the context of GUP, this parameter is deduced through the utilization of the Liouville theorem \cite{vagenas2019linear}. Consequently, the effective $\hbar$ is contingent on the momentum $p$ in the generalized uncertainty relation, resulting in a momentum-dependent size of the unit cell for each quantum state in phase space.

With such a modified partition function, one can easily obtained thermodynamic variables for the required statistics. We will mainly focus on pressure, number of particles, internal energy, and specific volume which are, respectively: 
\begin{align}
    P=&\,k_B T\frac{\partial}{\partial V} \mathrm{ln}Z, \label{therm1}\\
    n=&\,k_B T\frac{\partial}{\partial \mu} \mathrm{ln}Z\mid_{T,V}, \label{therm2}\\
    U=&\,k_B T^2\frac{\partial}{\partial T} \mathrm{ln}Z\mid_{z,V} \label{therm3}\\
    C_V=\,&\frac{\partial U}{\partial T}\mid_{V} \label{therm4}.
\end{align}

In what follows, we will predominantly center our attention on bosons, as some properties of Fermi particles in Palatini-like theories of gravity were studied in \cite{wojnar2023fermi,Wojnar:2023bvv}.

\section{Ideal Bose gas in the grand canonical ensemble }

Let us consider a simple system with $N$ identical spinless particles described by the non-interacting Hamiltonian
\begin{equation}
    H= \sum^N_{i=1} \frac{p_i^2}{2m},
\end{equation}
where $p_i^2=\mathrm{p}_i \cdot \mathrm{p}_i$ and $\mathrm{p}_i$ is the momentum operator of the single-particle with energy $E_p=p^2/2m$.
The grand partition function of an ideal Bose gas in the grand canonical ensemble for such a system is then given by \cite{huang2009introduction}:
\begin{equation}
    Z = \prod_p \frac{1}{1-z e^{-\beta E_p}},
\end{equation}
where $\beta =:(k_BT)^{-1}$. The equation of state is then expressed as
\begin{equation}\label{p1}
    \beta{PV}=-\sum_p \mathrm{ln}(1-ze^{-\beta E_p}),
\end{equation}
while the total number of particles
\begin{equation}\label{N1}
    N=z\frac{\partial}{\partial z}  \mathrm{ln} Z= \sum_p\frac{ze^{-\beta E_p}}{1-ze^{-\beta E_p}}.
\end{equation}
Since $N=\sum_{p}   \langle n_p \rangle$, an average occupation number for state $p$ is
\begin{equation}
    \langle n_p \rangle = -\frac{1}{\beta} \frac{\partial}{\partial E_p}\mathrm{ln}Z =\frac{ze^{-\beta E_p}}{1-ze^{-\beta E_p}}.
\end{equation}

The above series \eqref{p1} and \eqref{N1} are divergent for $z\rightarrow1$ because the term with $p=0$ diverges. Since the contribution of term with $p=0$ is important, let us then extract it while the rest of the series will be replaced by \eqref{sumint} as we consider $V\rightarrow\infty$: 
\begin{align}
   \beta P =& - \frac{4\pi}{(2\pi \hbar)^3} \int_0^\infty\frac{dp p^2}{1-\sigma p}
    \mathrm{ln}\left[1-z e^{-\beta\frac{p}{2m}}\right]- \frac{\mathrm{ln}(1-z)}{V} ,\\
    \frac{1}{v} =&  \frac{4\pi}{(2\pi \hbar)^3} \int_0^\infty\frac{dp p^2}{1-\sigma p}
    \frac{1}{z^{-1} e^{\beta\frac{p}{2m}}-1} + \frac{1}{V} \frac{z}{1-z},
\end{align}
where we have defined the specific volume $v=V/N$.

Introducing a new variable $x=\sqrt{\frac{\beta}{2m}}p$ with the thermal wavelength 
\begin{equation}
\lambda = \sqrt{\frac{2\pi\hbar^2}{mk_BT}},
\end{equation}
and considering a case when $|\alpha|=: |\sigma|\sqrt{2mk_BT}<\frac{1}{|x|}$ in the series expansion of the functions under the integrals to ensure that the results converge, the above expressions can be written as:
\begin{align}
   \beta P = -&\frac{4}{\pi}\frac{1}{\lambda^3} \int_0^\infty  \sum_{n=0}^\infty \alpha^n \left[ x^{2+n}\mathrm{ln}(1-ze^{-x^2})\right]  dx \nonumber \\
    -& \frac{\mathrm{ln}(1-z)}{V} ,\\
    \frac{1}{v} =& \frac{4}{\pi}\frac{1}{\lambda^3} \int_0^\infty  \sum_{n=0}^\infty  \frac{\alpha^n z x^{2+n}}{e^{x^2}-z} dx
     + \frac{1}{V} \frac{z}{1-z},
\end{align}

Since we are interested in the terms which are linear in $\alpha$, we can write
{ \footnotesize
\begin{align}
   \beta P 
   &= \frac{1}{\lambda^3}\left[ g_{5/2}(z) - \frac{4\alpha }{\pi} \int_0^\infty 
   \left(  
   x^3 \mathrm{ln}(1-ze^{-x^2}) \right)dx\right] \nonumber\\
  &  - \frac{\mathrm{ln}(1-z)}{V} + O(\alpha^2) ,\\
    \frac{1}{v} &= \frac{1}{\lambda^3} \left[ g_{3/2}(z) + \frac{4\alpha }{\pi} \int_0^\infty 
   \left(  
   \frac{x^3 z}{e^{x^2}-z} 
   \right) dx\right] \nonumber\\
    &+ \frac{1}{V} \frac{z}{1-z} + O(\alpha^2), \label{v}
\end{align}
}
where 
\begin{align}
    g_{5/2}(z) &= -\frac{4}{\pi} \int_0^\infty dx\, x^2 \mathrm{ln}(1-ze^{-x^2}) = \sum_{n=1}^\infty \frac{z^n}{n^{5/2}},\\
    g_{3/2}(z) &= z\frac{\partial}{\partial z}   g_{5/2}(z) = \sum_{n=1}^\infty \frac{z^n}{n^{3/2}}.
\end{align}
The above expressions can be further written in more compact forms as
\begin{align}
   \beta P 
   &= \frac{1}{\lambda^3}\left[ g_{5/2}(z) + \frac{2\alpha }{\pi}\mathrm{Li}_3(z)\right] - \frac{\mathrm{ln}(1-z)}{V} , \label{Eos}\\
    \frac{1}{v} &= \frac{1}{\lambda^3} \left[ g_{3/2}(z) + \frac{2\alpha }{\pi} \mathrm{Li}_2(z) \right] + \frac{1}{V} \frac{z}{1-z}, \label{v}
\end{align}
where $\mathrm{Li}_n(z)$ is the polylogarithm function and can be represented by a series of the form for $|z|<1$
\begin{equation}
  \mathrm{Li}_n(z) = \sum_{k=1}^{\infty} \frac{z^k}{k^n}  .
\end{equation}

Note that the last term in \eqref{v}
\begin{equation}\label{n0}
    \frac{z}{1-z} =   \langle n_0 \rangle
\end{equation}
is an occupation number for state $p=0$ and its contribution is large if $   \langle n_0 \rangle/V$ is a finite number. We will discuss the consequences of that in the next section.

The internal energy, since $\mathrm{ln}Z=\beta PV$, is also modified, taking the form 
\begin{equation}
   \frac{U}{V}  = -\frac{1}{V} \frac{\partial}{\partial \beta} \mathrm{ln} Z = 
    \frac{3k_BT}{2\lambda^3}\left[ g_{5/2}(z) + \frac{2\alpha }{\pi}\mathrm{Li}_3(z)\right] .
\end{equation}
However, notice that comparing it with \eqref{Eos} with the assumption that its last term can be neglected, we have a simple relation between the internal energy and temperature:
\begin{equation}
    U=\frac{3}{2}PV.
\end{equation}

Therefore, we have an equation of state given by \eqref{Eos} and \eqref{v} for the ideal Bose gas consisting of $N$ particles with mass $m$ contained in a vessel with a volume of $V$ in a framework of modified gravity and linear GUP models. To study its properties, we need to know the fugacity $z$ dependence on the temperature and specific volume $v$. To do so, let us now consider particular cases of the Bose gas in the framework of modified gravity and GUP.

\section{Bose-Einstein condensate}

\begin{figure*}[t]
\centering
\includegraphics[width=\textwidth]{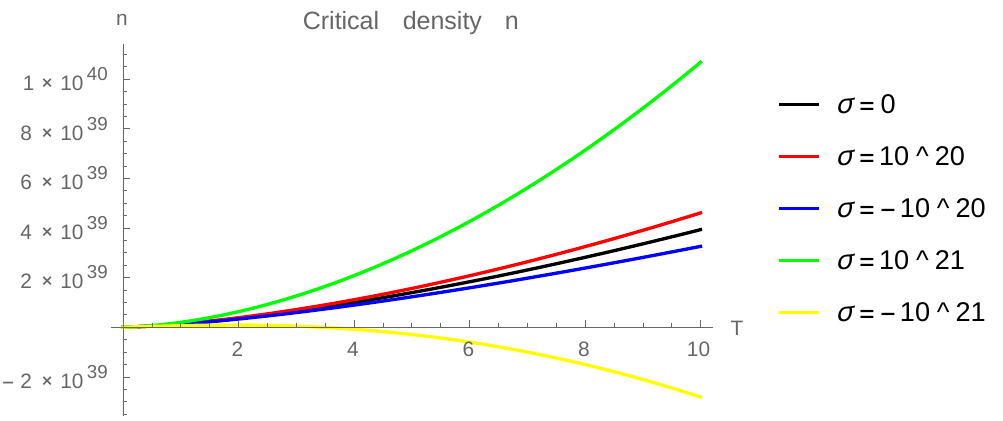}
\caption{[color online] Critical density (in m$^{-3}$), as described by Eq. \eqref{ncrit}, varies with temperature (in K) for several values of the parameter $\sigma$. It is crucial to observe that, to ensure a reasonable curve behavior, the parameter's value had to be reduced by approximately 2 orders of magnitude compared to the most recent bounds derived from seismic data \cite{Wojnar:2023bvv, Kozak:2023axy, Kozak:2023ruu}.
}
\label{fig1}
\end{figure*}

To examine the Bose-Einstein condensate in Ricci-based gravity, let us firstly analyze the behaviour of the fugacity $z$. To do so, we will focus on the equation \eqref{v}. It depends on the properties of the the functions $g_{3/2}(z)$ and $\mathrm{Li}_2(z)$. It results that the equation \eqref{v} has a solution only for $0\leq z \leq 1$. For $z=1$,
\begin{equation}
    g_{3/2}(1) = \zeta(3/2) \approx2.612,\;\;\;\mathrm{Li}_2(1) = \zeta(2)= \frac{\pi^2}{6}.
\end{equation}
Let us write \eqref{v} as 
\begin{align}
     \lambda^3 \frac{  \langle n_0 \rangle}{V} =     \frac{\lambda^3}{v} -
       \left[ g_{3/2}(z) + \frac{2\alpha }{\pi}\mathrm{Li}_2(z) \right].
\end{align}
We recognize the modified condition for the Bose-Einstein condensate: if temperature and specific volume satisfy (let us recall that $\alpha$ also depends on the temperature via its definition: $\alpha=: \sigma\sqrt{2mk_BT}$): 
\begin{equation}
    \frac{\lambda^3}{v} -\frac{\pi\alpha }{3} > g_{3/2}(1)  ,
\end{equation}
then $ \frac{  \langle n_0 \rangle}{V}$  is a finite number of all particles at the state with $p=0$. This condition defines a modified (with respect to the standard case with $\alpha=0)$ subspace of thermodynamic parameters $p$, $v$, $T$ of the ideal Bose gas where the Bose-Einstein condensate occurs. This region is separated from the rest of the $p$-$v$-$T$ space by the surface
\begin{align}
       \frac{\lambda^3}{v} - \frac{2\alpha }{\pi} \zeta(2)=
    g_{3/2}(1),
\end{align}
providing the critical value for the specific volume (or critical density $n_c=1/v_{cr}$) which clearly is modified by the gravity models:
\begin{equation}\label{ncrit}
       n_c= \left( \frac{1}{4\pi\hbar^2} \right)^\frac{3}{2} \left[ 
    \zeta(3/2) (2m k_B T)^\frac{3}{2} + \sigma\frac{\pi}{3} (2m k_B T)^2
    \right].
\end{equation}
In the non-deformed case $\sigma=0$ one can also easily determine the critical temperature as a function of $v_{cr}$. However, in modified gravity, one needs to deal with numerical solutions. However, similarly to the $\sigma=0$ case we can say that  we are dealing with a condensate when $T<T_{cr}$ (or $v<v_{cr}$) with slightly modified values. As we will see later, we can use that fact to put a bound on the theories' parameters. The plot of dependence of critical density \eqref{ncrit} on temperature and parameter $\sigma$ is given in Fig. \ref{fig1}.

The fugacity dependence on $T$ and $v$ is then
\[
 z = 
  \begin{cases} 
     1   \mbox{ for }\;\;\;\;\;\;\;\;\;  \frac{\lambda^3}{v} -\frac{\pi\alpha }{3} \geq g_{3/2}(1)& \\
    \mbox{solution of }   \frac{\lambda^3}{v} =
       \left[ g_{3/2}(z) + \frac{2\alpha }{\pi}\mathrm{Li}_2(z) \right]  & \mbox{otherwise. }
  \end{cases}
\]

Therefore, the fugacity is stuck at $1$ during the Einstein-Bose condensate, that is, the chemical potential is zero (that is, for $ \frac{\lambda^3}{v} -\frac{\pi\alpha }{3} \leq g_{3/2}(1)$ region we deal with the gas phase).

Let us now write the equation of state and other thermodynamic functions in both regions:
\[
 \beta P =
  \begin{cases} 
  \frac{1}{\lambda^3}\left[ g_{5/2}(z) + \frac{2\alpha }{\pi}\mathrm{Li}_3(z)\right]  & \mbox{if } v> v_{cr},  \\
   \frac{1}{\lambda^3}\left[ g_{5/2}(1) + \frac{2\alpha }{\pi}\zeta(3) \right]  & \mbox{if } v< v_{cr},
  \end{cases}
\]

\[
 \frac{U}{N} =\frac{3}{2}Pv=
  \begin{cases} 
\frac{3}{2}  \frac{k_BTv}{\lambda^3}\left[ g_{5/2}(z) + \frac{2\alpha }{\pi}\mathrm{Li}_3(z)\right]  & \mbox{if } v> v_{cr},  \\
 \frac{3}{2}  \frac{k_BTv}{\lambda^3}\left[ g_{5/2}(1) + \frac{2\alpha }{\pi}\zeta(3) \right]  & \mbox{if } v< v_{cr},
  \end{cases}
\]

\[
 \frac{C_V}{Nk_B} =
  \begin{cases} 
\frac{15}{4}\frac{v}{\lambda^3} h_1(T) + \frac{3}{2}\frac{Tv}{\lambda^3} 
h_2(T) \frac{dz}{dT}
  & \mbox{if } v> v_{cr},  \\
\frac{15}{4}\frac{v}{\lambda^3}g_{5/2}(1)+\sigma f_1(T)  & \mbox{if } v< v_{cr},
  \end{cases}
\]
where 
\begin{align}
h_1(T) &= g_{5/2}(z)+\frac{14\sigma\sqrt{2mk_BT}}{\pi}\mathrm{Li}_3(z),\\
h_2(T) &= \frac{g_{3/2}(z)}{z} + \frac{2\sigma\sqrt{2mk_BT}}{\pi}\frac{\mathrm{Li}_2(z)}{z},\\
    f_1(T)&=\frac{3v}{4\lambda^3} \left(
\frac{10\zeta(3)}{\pi}\sqrt{2mk_BT}+\frac{mk_BT}{\lambda}\frac{\zeta(3)}{\pi^3\hbar^4}
    \right).
\end{align}
The derivative $(dz/dT)_V$ is also modified and is given as
\begin{align}
    \frac{dz}{dT} = &-\left(
    \frac{3\lambda^3}{2vT}+ \frac{\sigma}{\pi}\sqrt{\frac{2mk_B}{T}}\mathrm{Li}_2(z)
    \right) \nonumber \\
  &  \times
    \left(
    \frac{g_{1/2}(z)}{z} + \frac{2\sigma \sqrt{2mk_B T}}{\pi} \frac{\mathrm{Li}_1(z)}{z}.
    \right)^{-1}
\end{align}

\begin{figure*}[t]
\centering
\includegraphics[width=\textwidth]{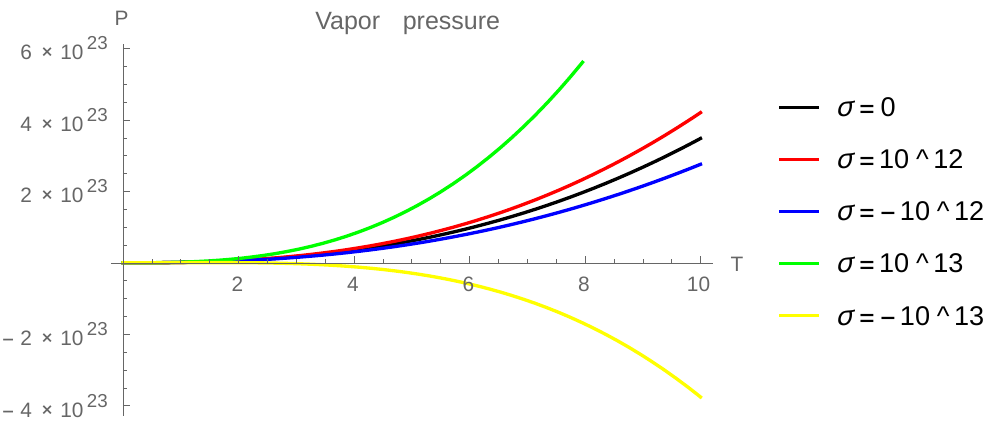}
\caption{[color online] Vapor pressure (in Pa), expressed by Eq. \eqref{vapor}, varies with temperature (in K) for several $\sigma$ values. It is important to highlight that to ensure a sensible curve behavior, we had to reduce the parameter's value by approximately 10 orders of magnitude compared to the latest bounds derived from seismic data \cite{Wojnar:2023bvv, Kozak:2023axy, Kozak:2023ruu}.}
\label{fig2}
\end{figure*}

The vapor pressure is then given by the expression 
\begin{equation}\label{vapor}
    P_0(T) =   \frac{k_BT}{\lambda^3}\left[ g_{5/2}(1) + \frac{2\sigma\sqrt{2mk_BT} }{\pi}\zeta(3) \right]
\end{equation}
while its plot is given in Fig. \ref{fig2}.
The derivative of the vapor pressure with respect to the temperature is
\begin{equation}\label{prim}
    \frac{dP_0(T)}{dT} = \frac{5}{2} \frac{k_B g_{5/2}(1)}{g_{3/2}(1)v_{cr}} \left( 1+\frac{\sigma}{\pi} f_2(T)\right) ,
\end{equation}
where
\begin{equation}
    f_2(T) = \frac{48\sqrt{\frac{\hbar^2\pi}{k_B}}}{5g_{5/2}(1)(g_{3/2}(1)v_{cr})^{1/3}}\zeta(3)
    -\frac{\sqrt{8k_B m T}\zeta(2)}{g_{3/2}(1)} .
\end{equation}
Writing \eqref{prim} as
\begin{equation}
     \frac{dP_0(T)}{dT} = \frac{1}{Tv_{cr}}\left[\frac{5}{2} \frac{k_B T g_{5/2}(1)}{g_{3/2}(1)} \left( 1+\frac{\sigma}{\pi} f_2(T)\right) \right]
\end{equation}
we see that we deal with the Clapeyron equation with a modified latent heat having the following form
\begin{equation}\label{latent}
    L=\frac{5}{2} \frac{k_B T g_{5/2}(1)}{g_{3/2}(1)} \left( 1+\frac{\sigma}{\pi} f_2(T)\right).
\end{equation}
Therefore, in GUP models and modified gravity, the Bose-Einstein condensation is also a first-order phase transition if $L\neq0$. We will come back to that issue in the end of this section.

Let us now come back to the critical density \eqref{ncrit}. Assuming that He$^{4}$ is an ideal Bose gas in the condensate state and applying the experimental data related to the transition point
\begin{equation}
    T_c= 2.172\,\text{ K},\,\,\,\,n_c=2.16\times10^{28}\text{ m}^{-3},
\end{equation}
we can find the deformed parameter corresponding to those values:
\begin{equation}
    \sigma\approx2.837\times10^{24}\frac{ \text{s}}{\text{kg m}},
\end{equation}
providing, that the Palatini parameter is
\begin{equation}
    \bar\beta \approx 9.352\times 10^{10}\text{ m}^{2}.
\end{equation}
On the other hand, inserting the same critical values for the temperature and specific volume ($v_{cr}=n_c^{-1}$) in the bracket expression in \eqref{latent} we obtain that the latent heat vanishes for $\sigma\approx-3.6\times 10^{12}$. It also explain the non-physical vapor pressure's behavior in the Fig. \ref{fig2} for higher orders of magnitude for the negative values of the deformation parameter.

Nevertheless, using the idealization such as the non-interacting Bose gas for explaining the behaviour of the liquid helium in low temperature does not provide us improved bounds for the parameters introduced by quantum and modified gravity. The order or magnitude for the upper bound is about 2 times worse than in the case of the recently developed methodology in which one uses the Earth's seismic data \cite{Kozak:2023axy,Kozak:2023ruu,Wojnar:2023bvv}. However, we are aware that in the case of He$^4$ one deals with a second order phase transition which is interpreted as Bose-Einstein condensate with the strong interatomic interactions taken into account. Considering more realistic models one expects to obtain better constraints. Because of that fact, we will now focus on the Landau model for liquid helium which was proved to provide a reasonable description of the He$^4$ behaviour in low temperatures.

\section{liquid helium }

The Landau model \cite{landau2018theory,tisza1947theory} provides a comprehensive microscopic description of a two-fluid model near absolute zero. The specific heat of liquid helium as $T\rightarrow0$ behaves as $T^3$ (note that in the ideal Bose gas, we have $C_v\sim T^{3/2}$, as discussed in the previous section), which is characteristic of a phonon gas and has been experimentally confirmed. On the other hand, in the finite-temperature regime, an additional term comes into play. Thus, the energy (dispersion relation) of quasiparticles as a function of wave number $k$ for He$^4$ can be expressed as:
\[
 \hbar \omega =
  \begin{cases} 
 \hbar ck & \mbox{if } k<< k_0,  \\
 \Delta+ \frac{\hbar^2(k-k_0)^2}{2\gamma}  & \mbox{if } k\approx k_0,
  \end{cases}
\]
where $c$ is the sound velocity while $\Delta$, $k_0$, $\gamma$ are experimental constants. In the Landau theory, one assumes that the quantum states of He$^4$ close to the ground state can be considered as the states of a non-interacting gas with energy levels
\begin{align}\label{uint}
    U=&E_0+\sum_k \hbar \omega_k \langle n_k \rangle \nonumber \\
    =& E_0 + \frac{V}{2\pi^2} \int^\infty_0 \frac{k^2\hbar\omega_k}{e^{\beta\hbar\omega_k}-1}\frac{dk}{(1-\sigma\hbar k)}.
\end{align}
Here, $ \hbar \omega_k $ represents the elementary excitation energy with the wave vector $k$ and occupation number $\langle n_k \rangle$. In the second equality, we have already considered the deformation of the phase space. Now, let us calculate the internal energy and its GUP corrections at low temperatures. In this scenario, only the contributions from the phonon and roton parts \cite{cohen1957theory,yarnell1959excitations} contribute to the energy in Eq. \eqref{uint}. The phonon part is expressed as:
\begin{equation}\label{ephon}
    E_\text{phonon}=\frac{V}{2\pi^2}\left( \frac{\pi^4(k_BT)^4}{15\hbar^3 c^3} + 24\sigma \frac{(k_BT)^5\zeta(5)}{c^4\hbar^3}\right).
\end{equation}
Therefore, the phonon specific heat is given by
\begin{equation}
    \frac{C_V^\text{phonon}}{k_BN} = \frac{2\pi^2v(k_BT)^3}{15\hbar^3c^3}+
    60\sigma \frac{\zeta(5)(k_B T)^4v}{c^4\hbar^3\pi^2}.
\end{equation}
We can calculate the roton part assuming that $\beta\Delta$ is small. This provides that the energy is
\begin{equation}\label{eroton}
    \frac{E_\text{roton}}{V}\approx \frac{k_0^2\Delta}{\pi}\sqrt{\frac{\gamma k_BT}{2\pi\hbar^2}}e^{-\frac{\Delta}{k_BT}}(1+\sigma \hbar k_0),
\end{equation}
providing that the roton specific heat
\begin{equation}
    \frac{C_V^\text{roton}}{k_BN} = \frac{k_0^2v\Delta^2}{\pi}\sqrt{\frac{\gamma k_BT}{2\pi\hbar^2}}\frac{e^{-\frac{\Delta}{k_BT}}}{(k_BT)^2}(1+\sigma \hbar k_0).
\end{equation}
Note that we can also obtain values of the parameter $\sigma$ for which phonon \eqref{ephon} and roton \eqref{eroton} energies vanishes. This happens for $\sigma\approx-10^{24}$ ($\bar\beta\approx-5\times10^{10}$) and $\sigma\approx-10^{23}$ ($\bar\beta\approx-10^{10}$), respectively.

\begin{figure*}[t]
\centering
\includegraphics[width=\textwidth]{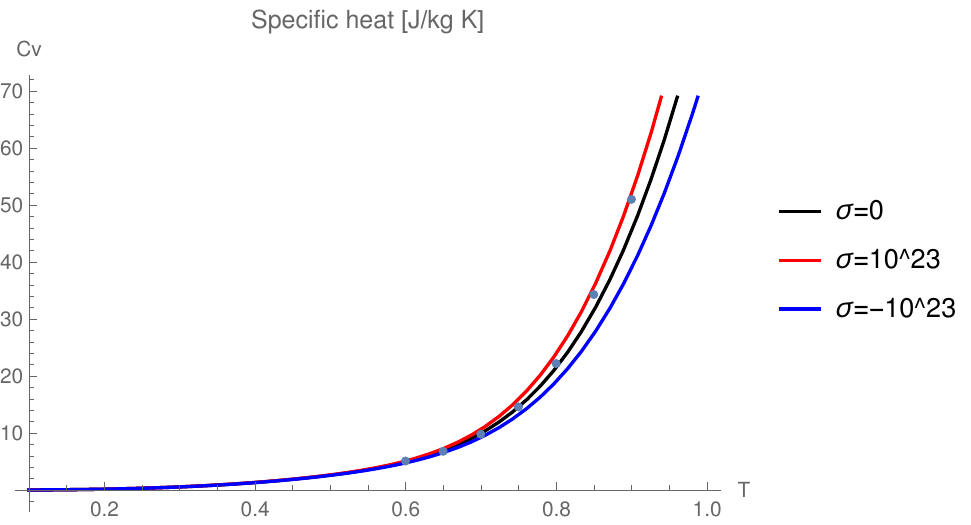}
\caption{[color online] Specific heat of He$^{4}$ as a function of temperature and deformation parameter $\sigma$ given by the Eq. \eqref{sheat}. { Therefore, the resulting constraints (see the text) allows us to constrain linear GUP and Ricci-based gravities models.} The data points are taken from \cite{kramers1957chapter}.
}
\label{fig3}
\end{figure*}

The specific heat for liquid helium in low temperature is then a sum of those two specific heats. Applying the numerical values for the experimental data \cite{yarnell1959excitations} (note that $Nv=\rho^{-1}$)
\begin{align*}
    c=239\text{ m s}^{-1},\,\,\rho=144\text{ kg m}^{-3},\,\,\Delta/k_B=8.65\text{K},\\
    k_0 =
    1.92\times10^{10}\text{ m}^{-1},\,\,\gamma = 1.07\times10^{-27}\text{ kg}
\end{align*}
we have (in Jkg$^{-1}$K$^{-1}$) 
\begin{align}\label{sheat}
    C_{\text{He}^4}&=20.7 T^3+ \frac{387\times10^{3}}{T^{3/2}} e^{-8.85/T} \\
    &+\sigma (5.73\times10^{-24}T^4 +\frac{7.83\times10^{-19}}{T^{3/2}} e^{-8.85/T}). \nonumber
\end{align}
The plot of specific heat of He$^4$ as a function of temperature is given in Fig. \ref{fig3}. We have also plotted the data points from \cite{kramers1957chapter}. The discrepancy for $\sigma=0$ when $T\rightarrow1$ K is believed to arise due to the approximation taken in the roton part of the specific heat, while one has a good fit up to $T\approx 0.8$ K. We see that quantum and modified gravity corrections to both heats (with a similar approximation performed as for the case $\sigma=0$) provide better fit to the data for the parameter $\sigma$ of the order $10^{23}$. To be more specific, if we consider for instance, the data point $(T=0.6,\,C_{\text{He}^4}=5.1)$, we obtain the value of the deformation parameter $\sigma=1.04\times10^{23}$ ($\bar\beta=9.4\times10^{10}$).

\section{Conclusions}

The aim of this paper was to study the effects of Ricci-based gravity, such as Palatini $f(R)$ and Eddington-inspired Born-Infel models, and linear Generalized Uncertainty Principle, on the systems described by the Bose statistics. Using the recently derived correspondence between modified gravity and GUP models, we were able to provide the formalism allowing to study the ideal Bose gases. As expected, modified gravity (or linear GUP) introduces additional terms to the well-known expressions, allowing us to constrain the gravitational models with the tabletop experiments. 

As working examples, we firstly analyzed the Bose-Einstein condensate. The effects of the phase space deformation modifies slightly the critical values which are the boundary values for the condensate to happen. For instant, in order not to differ too much from the non-deformed case, the deformation parameter $\sigma$ would have to be about 2 orders of magnitude smaller than the bounds given by the methodology in which the Earth seismic data were used \cite{Kozak:2023axy,Kozak:2023ruu,Wojnar:2023bvv}. On the other hand, using the similar arguments, the derivation of the vapor pressure (which also depicts the transition line) reveals that the order of the bounds should be about 10 order less in order not to change the microscopic behaviour of the gas too much.

We have also obtained that the Bose-Einstein condensate is the first-order transition with a modified latent heat. Interestingly, there exists such a value of the deformation parameter $\sigma$ (or the Palatini parameter $\bar\beta$) for which the latent heat vanishes. Such a singular value (that is, it is related to the spacetime curvature or/and minimal length), depends on the temperature and specific volume, and can be responsible for a kind of the phase transition which we deal with. For the critical temperature and density its value is
$$\sigma\approx-3.6\times 10^{12}\frac{ \text{s}}{\text{kg m}},$$
for the deformation parameter of the linear GUP, while the Palatini parameter is
$$
    \bar\beta \approx -0.12\text{ m}^{2},
$$
which we can consider as lower bounds arising from the ideal Bose-Einstein condensate.

However, assuming that the liquid helium He$^4$ is the ideal Bose gas with the temperature and critical density provided by the experiments, we obtained worse bounds that the ones provided by seismology, that is,
$$
    \sigma\approx2.837\times10^{24}\frac{ \text{s}}{\text{kg m}},
$$
for the deformation parameter of the linear GUP, while the Palatini parameter is
$$
    \bar\beta \approx 9.352\times 10^{10}\text{ m}^{2}.
$$
Those are the upper bounds resulting from the analysis of the ideal Bose-Einstein condensate.

In order to have an inside into a more realistic description of the behaviour of liquid helium in low temperatures, we have also analysed the Landau model. Deriving the curvature corrections to the specific heat of phonons and rotons, we could compare our theoretical results with the experimental ones and get an idea about the order of magnitude of the acceptable values of the deformation parameter. It is $$\sigma\approx10^{23}\frac{ \text{s}}{\text{kg m}}$$
for the linear GUP models and $$\bar\beta\approx3\times10^{9}\text{ m}^{2}$$ for Palatini gravity ($\epsilon\approx1.2\times10^{10}\text{ m}^{2}$ for EiBI). We regard it as an upper bound for the parameters, albeit less stringent than the one derived from seismic data. However, by incorporating a more accurate depiction of interatomic interactions and the nature of excited states, such as in the Feynman model of He$^4$ \cite{cohen1957theory}, alongside the latest data, we anticipate enhancing the current constraints on quantum and modified gravity parameters.

Furthermore, similar to the ideal Bose gas, we can identify specific values of this parameter where the phonon and roton specific heats vanish in the low-temperature regime. This occurs around $\sigma\approx-10^{24}$ for the phonon part and $\sigma\approx-10^{23}$ for the roton part. This order of magnitude is more realistic compared to the ideal Bose-Einstein condensate case, as it pertains to a physical system whose behavior is validated by tabletop experiments (with very low but not zero temperature the specific heats are not zeros). Consequently, we view this as a lower bound for the deformation parameter.

Upon comparing our results, derived from tabletop experiments, with those obtained using a novel method employing Earth's seismic data, the latter case yields a more constrained range for the parameters.

In this paper, an examination of an idealized case involving Bose-Einstein condensate and a realistic description of the behavior of liquid helium at low temperatures illustrates the invaluable utility of the connection between Modified Gravity and Generalized Uncertainty Principle models. This correspondence serves as extremely useful tools for studying various physical systems at different scales. This not only facilitates constraining different proposals related to quantum and modified gravity but also enables the analysis of gravitational effects through tabletop experiments. Ongoing research along these lines seeks to delve deeper into and validate the implications of Modified Gravity on the microscopic properties of matter.

\section*{Acknowledgements}
 AW acknowledges financial support from MICINN (Spain) {\it Ayuda Juan de la Cierva - incorporaci\'on} 2020 No. IJC2020-044751-I.

\bibliographystyle{apsrev4-1}
\bibliography{biblio}

\begin{thebibliography}{73}%
\makeatletter
\providecommand \@ifxundefined [1]{%
 \@ifx{#1\undefined}
}%
\providecommand \@ifnum [1]{%
 \ifnum #1\expandafter \@firstoftwo
 \else \expandafter \@secondoftwo
 \fi
}%
\providecommand \@ifx [1]{%
 \ifx #1\expandafter \@firstoftwo
 \else \expandafter \@secondoftwo
 \fi
}%
\providecommand \natexlab [1]{#1}%
\providecommand \enquote  [1]{``#1''}%
\providecommand \bibnamefont  [1]{#1}%
\providecommand \bibfnamefont [1]{#1}%
\providecommand \citenamefont [1]{#1}%
\providecommand \href@noop [0]{\@secondoftwo}%
\providecommand \href [0]{\begingroup \@sanitize@url \@href}%
\providecommand \@href[1]{\@@startlink{#1}\@@href}%
\providecommand \@@href[1]{\endgroup#1\@@endlink}%
\providecommand \@sanitize@url [0]{\catcode `\\12\catcode `\$12\catcode `\&12\catcode `\#12\catcode `\^12\catcode `\_12\catcode `\%12\relax}%
\providecommand \@@startlink[1]{}%
\providecommand \@@endlink[0]{}%
\providecommand \url  [0]{\begingroup\@sanitize@url \@url }%
\providecommand \@url [1]{\endgroup\@href {#1}{\urlprefix }}%
\providecommand \urlprefix  [0]{URL }%
\providecommand \Eprint [0]{\href }%
\providecommand \doibase [0]{http://dx.doi.org/}%
\providecommand \selectlanguage [0]{\@gobble}%
\providecommand \bibinfo  [0]{\@secondoftwo}%
\providecommand \bibfield  [0]{\@secondoftwo}%
\providecommand \translation [1]{[#1]}%
\providecommand \BibitemOpen [0]{}%
\providecommand \bibitemStop [0]{}%
\providecommand \bibitemNoStop [0]{.\EOS\space}%
\providecommand \EOS [0]{\spacefactor3000\relax}%
\providecommand \BibitemShut  [1]{\csname bibitem#1\endcsname}%
\let\auto@bib@innerbib\@empty
\bibitem [{\citenamefont {Abbott}\ \emph {et~al.}(2016)\citenamefont {Abbott}, \citenamefont {Abbott}, \citenamefont {Abbott}, \citenamefont {Abernathy}, \citenamefont {Acernese}, \citenamefont {Ackley}, \citenamefont {Adams}, \citenamefont {Adams}, \citenamefont {Addesso}, \citenamefont {Adhikari} \emph {et~al.}}]{abbott2016observation}%
  \BibitemOpen
  \bibfield  {author} {\bibinfo {author} {\bibfnamefont {B.~P.}\ \bibnamefont {Abbott}}, \bibinfo {author} {\bibfnamefont {R.}~\bibnamefont {Abbott}}, \bibinfo {author} {\bibfnamefont {T.}~\bibnamefont {Abbott}}, \bibinfo {author} {\bibfnamefont {M.}~\bibnamefont {Abernathy}}, \bibinfo {author} {\bibfnamefont {F.}~\bibnamefont {Acernese}}, \bibinfo {author} {\bibfnamefont {K.}~\bibnamefont {Ackley}}, \bibinfo {author} {\bibfnamefont {C.}~\bibnamefont {Adams}}, \bibinfo {author} {\bibfnamefont {T.}~\bibnamefont {Adams}}, \bibinfo {author} {\bibfnamefont {P.}~\bibnamefont {Addesso}}, \bibinfo {author} {\bibfnamefont {R.}~\bibnamefont {Adhikari}},  \emph {et~al.},\ }\href@noop {} {\bibfield  {journal} {\bibinfo  {journal} {Physical review letters}\ }\textbf {\bibinfo {volume} {116}},\ \bibinfo {pages} {061102} (\bibinfo {year} {2016})}\BibitemShut {NoStop}%
\bibitem [{\citenamefont {Abbott}\ \emph {et~al.}(2017)\citenamefont {Abbott}, \citenamefont {Abbott}, \citenamefont {Abbott}, \citenamefont {Acernese}, \citenamefont {Ackley}, \citenamefont {Adams}, \citenamefont {Adams}, \citenamefont {Addesso}, \citenamefont {Adhikari}, \citenamefont {Adya} \emph {et~al.}}]{abbott2017gw170817}%
  \BibitemOpen
  \bibfield  {author} {\bibinfo {author} {\bibfnamefont {B.~P.}\ \bibnamefont {Abbott}}, \bibinfo {author} {\bibfnamefont {R.}~\bibnamefont {Abbott}}, \bibinfo {author} {\bibfnamefont {T.}~\bibnamefont {Abbott}}, \bibinfo {author} {\bibfnamefont {F.}~\bibnamefont {Acernese}}, \bibinfo {author} {\bibfnamefont {K.}~\bibnamefont {Ackley}}, \bibinfo {author} {\bibfnamefont {C.}~\bibnamefont {Adams}}, \bibinfo {author} {\bibfnamefont {T.}~\bibnamefont {Adams}}, \bibinfo {author} {\bibfnamefont {P.}~\bibnamefont {Addesso}}, \bibinfo {author} {\bibfnamefont {R.}~\bibnamefont {Adhikari}}, \bibinfo {author} {\bibfnamefont {V.~B.}\ \bibnamefont {Adya}},  \emph {et~al.},\ }\href@noop {} {\bibfield  {journal} {\bibinfo  {journal} {Physical review letters}\ }\textbf {\bibinfo {volume} {119}},\ \bibinfo {pages} {161101} (\bibinfo {year} {2017})}\BibitemShut {NoStop}%
\bibitem [{\citenamefont {Rubin}\ \emph {et~al.}(1980)\citenamefont {Rubin}, \citenamefont {Ford~Jr},\ and\ \citenamefont {Thonnard}}]{rubin1980rotational}%
  \BibitemOpen
  \bibfield  {author} {\bibinfo {author} {\bibfnamefont {V.~C.}\ \bibnamefont {Rubin}}, \bibinfo {author} {\bibfnamefont {W.~K.}\ \bibnamefont {Ford~Jr}}, \ and\ \bibinfo {author} {\bibfnamefont {N.}~\bibnamefont {Thonnard}},\ }\href@noop {} {\bibfield  {journal} {\bibinfo  {journal} {Astrophysical Journal, Part 1, vol. 238, June 1, 1980, p. 471-487.}\ }\textbf {\bibinfo {volume} {238}},\ \bibinfo {pages} {471} (\bibinfo {year} {1980})}\BibitemShut {NoStop}%
\bibitem [{\citenamefont {Huterer}\ and\ \citenamefont {Turner}(1999)}]{huterer1999prospects}%
  \BibitemOpen
  \bibfield  {author} {\bibinfo {author} {\bibfnamefont {D.}~\bibnamefont {Huterer}}\ and\ \bibinfo {author} {\bibfnamefont {M.~S.}\ \bibnamefont {Turner}},\ }\href@noop {} {\bibfield  {journal} {\bibinfo  {journal} {Physical Review D}\ }\textbf {\bibinfo {volume} {60}},\ \bibinfo {pages} {081301} (\bibinfo {year} {1999})}\BibitemShut {NoStop}%
\bibitem [{\citenamefont {Copeland}\ \emph {et~al.}(2006)\citenamefont {Copeland}, \citenamefont {Sami},\ and\ \citenamefont {Tsujikawa}}]{copeland2006dynamics}%
  \BibitemOpen
  \bibfield  {author} {\bibinfo {author} {\bibfnamefont {E.~J.}\ \bibnamefont {Copeland}}, \bibinfo {author} {\bibfnamefont {M.}~\bibnamefont {Sami}}, \ and\ \bibinfo {author} {\bibfnamefont {S.}~\bibnamefont {Tsujikawa}},\ }\href@noop {} {\bibfield  {journal} {\bibinfo  {journal} {International Journal of Modern Physics D}\ }\textbf {\bibinfo {volume} {15}},\ \bibinfo {pages} {1753} (\bibinfo {year} {2006})}\BibitemShut {NoStop}%
\bibitem [{\citenamefont {Nojiri}\ and\ \citenamefont {Odintsov}(2007)}]{nojiri2007introduction}%
  \BibitemOpen
  \bibfield  {author} {\bibinfo {author} {\bibfnamefont {S.}~\bibnamefont {Nojiri}}\ and\ \bibinfo {author} {\bibfnamefont {S.~D.}\ \bibnamefont {Odintsov}},\ }\href@noop {} {\bibfield  {journal} {\bibinfo  {journal} {International Journal of Geometric Methods in Modern Physics}\ }\textbf {\bibinfo {volume} {4}},\ \bibinfo {pages} {115} (\bibinfo {year} {2007})}\BibitemShut {NoStop}%
\bibitem [{\citenamefont {Baker}\ \emph {et~al.}(2015)\citenamefont {Baker}, \citenamefont {Psaltis},\ and\ \citenamefont {Skordis}}]{baker2015linking}%
  \BibitemOpen
  \bibfield  {author} {\bibinfo {author} {\bibfnamefont {T.}~\bibnamefont {Baker}}, \bibinfo {author} {\bibfnamefont {D.}~\bibnamefont {Psaltis}}, \ and\ \bibinfo {author} {\bibfnamefont {C.}~\bibnamefont {Skordis}},\ }\href@noop {} {\bibfield  {journal} {\bibinfo  {journal} {The Astrophysical Journal}\ }\textbf {\bibinfo {volume} {802}},\ \bibinfo {pages} {63} (\bibinfo {year} {2015})}\BibitemShut {NoStop}%
\bibitem [{\citenamefont {Kulikov}\ and\ \citenamefont {Pronin}(1995)}]{kulikov1995low}%
  \BibitemOpen
  \bibfield  {author} {\bibinfo {author} {\bibfnamefont {I.~K.}\ \bibnamefont {Kulikov}}\ and\ \bibinfo {author} {\bibfnamefont {P.~I.}\ \bibnamefont {Pronin}},\ }\href@noop {} {\bibfield  {journal} {\bibinfo  {journal} {International Journal of Theoretical Physics}\ }\textbf {\bibinfo {volume} {34}},\ \bibinfo {pages} {1843} (\bibinfo {year} {1995})}\BibitemShut {NoStop}%
\bibitem [{\citenamefont {Kim}(2014)}]{kim2014physics}%
  \BibitemOpen
  \bibfield  {author} {\bibinfo {author} {\bibfnamefont {H.-C.}\ \bibnamefont {Kim}},\ }\href@noop {} {\bibfield  {journal} {\bibinfo  {journal} {Physical Review D}\ }\textbf {\bibinfo {volume} {89}},\ \bibinfo {pages} {064001} (\bibinfo {year} {2014})}\BibitemShut {NoStop}%
\bibitem [{\citenamefont {Sakstein}(2015{\natexlab{a}})}]{sakstein2015testing}%
  \BibitemOpen
  \bibfield  {author} {\bibinfo {author} {\bibfnamefont {J.}~\bibnamefont {Sakstein}},\ }\href@noop {} {\bibfield  {journal} {\bibinfo  {journal} {Physical Review D}\ }\textbf {\bibinfo {volume} {92}},\ \bibinfo {pages} {124045} (\bibinfo {year} {2015}{\natexlab{a}})}\BibitemShut {NoStop}%
\bibitem [{\citenamefont {Kalita}\ \emph {et~al.}(2023)\citenamefont {Kalita}, \citenamefont {Sarmah},\ and\ \citenamefont {Wojnar}}]{Kalita:2022trq}%
  \BibitemOpen
  \bibfield  {author} {\bibinfo {author} {\bibfnamefont {S.}~\bibnamefont {Kalita}}, \bibinfo {author} {\bibfnamefont {L.}~\bibnamefont {Sarmah}}, \ and\ \bibinfo {author} {\bibfnamefont {A.}~\bibnamefont {Wojnar}},\ }\href {\doibase 10.1103/PhysRevD.107.044072} {\bibfield  {journal} {\bibinfo  {journal} {Phys. Rev. D}\ }\textbf {\bibinfo {volume} {107}},\ \bibinfo {pages} {044072} (\bibinfo {year} {2023})},\ \Eprint {http://arxiv.org/abs/2212.04918} {arXiv:2212.04918 [gr-qc]} \BibitemShut {NoStop}%
\bibitem [{\citenamefont {Wojnar}\ and\ \citenamefont {Velten}(2016)}]{Wojnar:2016bzk}%
  \BibitemOpen
  \bibfield  {author} {\bibinfo {author} {\bibfnamefont {A.}~\bibnamefont {Wojnar}}\ and\ \bibinfo {author} {\bibfnamefont {H.}~\bibnamefont {Velten}},\ }\href {\doibase 10.1140/epjc/s10052-016-4549-z} {\bibfield  {journal} {\bibinfo  {journal} {Eur. Phys. J. C}\ }\textbf {\bibinfo {volume} {76}},\ \bibinfo {pages} {697} (\bibinfo {year} {2016})},\ \Eprint {http://arxiv.org/abs/1604.04257} {arXiv:1604.04257 [gr-qc]} \BibitemShut {NoStop}%
\bibitem [{\citenamefont {Wojnar}(2018)}]{Wojnar:2017tmy}%
  \BibitemOpen
  \bibfield  {author} {\bibinfo {author} {\bibfnamefont {A.}~\bibnamefont {Wojnar}},\ }\href {\doibase 10.1140/epjc/s10052-018-5900-3} {\bibfield  {journal} {\bibinfo  {journal} {Eur. Phys. J. C}\ }\textbf {\bibinfo {volume} {78}},\ \bibinfo {pages} {421} (\bibinfo {year} {2018})},\ \Eprint {http://arxiv.org/abs/1712.01943} {arXiv:1712.01943 [gr-qc]} \BibitemShut {NoStop}%
\bibitem [{\citenamefont {Sarmah}\ \emph {et~al.}(2022)\citenamefont {Sarmah}, \citenamefont {Kalita},\ and\ \citenamefont {Wojnar}}]{Sarmah:2021ule}%
  \BibitemOpen
  \bibfield  {author} {\bibinfo {author} {\bibfnamefont {L.}~\bibnamefont {Sarmah}}, \bibinfo {author} {\bibfnamefont {S.}~\bibnamefont {Kalita}}, \ and\ \bibinfo {author} {\bibfnamefont {A.}~\bibnamefont {Wojnar}},\ }\href {\doibase 10.1103/PhysRevD.105.024028} {\bibfield  {journal} {\bibinfo  {journal} {Phys. Rev. D}\ }\textbf {\bibinfo {volume} {105}},\ \bibinfo {pages} {024028} (\bibinfo {year} {2022})},\ \Eprint {http://arxiv.org/abs/2111.08029} {arXiv:2111.08029 [gr-qc]} \BibitemShut {NoStop}%
\bibitem [{\citenamefont {Wojnar}(2020)}]{Wojnar:2020txr}%
  \BibitemOpen
  \bibfield  {author} {\bibinfo {author} {\bibfnamefont {A.}~\bibnamefont {Wojnar}},\ }\href {\doibase 10.1103/PhysRevD.102.124045} {\bibfield  {journal} {\bibinfo  {journal} {Phys. Rev. D}\ }\textbf {\bibinfo {volume} {102}},\ \bibinfo {pages} {124045} (\bibinfo {year} {2020})},\ \Eprint {http://arxiv.org/abs/2007.13451} {arXiv:2007.13451 [gr-qc]} \BibitemShut {NoStop}%
\bibitem [{\citenamefont {Guerrero}\ \emph {et~al.}(2021)\citenamefont {Guerrero}, \citenamefont {Rubiera-Garcia},\ and\ \citenamefont {Wojnar}}]{Guerrero:2021fnz}%
  \BibitemOpen
  \bibfield  {author} {\bibinfo {author} {\bibfnamefont {M.}~\bibnamefont {Guerrero}}, \bibinfo {author} {\bibfnamefont {D.}~\bibnamefont {Rubiera-Garcia}}, \ and\ \bibinfo {author} {\bibfnamefont {A.}~\bibnamefont {Wojnar}},\ }\href@noop {} {\  (\bibinfo {year} {2021})},\ \Eprint {http://arxiv.org/abs/2112.03682} {arXiv:2112.03682 [gr-qc]} \BibitemShut {NoStop}%
\bibitem [{\citenamefont {Gomes}\ and\ \citenamefont {Wojnar}(2022)}]{Gomes:2022sft}%
  \BibitemOpen
  \bibfield  {author} {\bibinfo {author} {\bibfnamefont {D.~A.}\ \bibnamefont {Gomes}}\ and\ \bibinfo {author} {\bibfnamefont {A.}~\bibnamefont {Wojnar}},\ }\href@noop {} {\  (\bibinfo {year} {2022})},\ \Eprint {http://arxiv.org/abs/2206.04464} {arXiv:2206.04464 [gr-qc]} \BibitemShut {NoStop}%
\bibitem [{\citenamefont {Wojnar}(2023{\natexlab{a}})}]{wojnar2023fermi}%
  \BibitemOpen
  \bibfield  {author} {\bibinfo {author} {\bibfnamefont {A.}~\bibnamefont {Wojnar}},\ }\href@noop {} {\bibfield  {journal} {\bibinfo  {journal} {Physical Review D}\ }\textbf {\bibinfo {volume} {107}},\ \bibinfo {pages} {044025} (\bibinfo {year} {2023}{\natexlab{a}})}\BibitemShut {NoStop}%
\bibitem [{\citenamefont {Sakstein}(2015{\natexlab{b}})}]{sakstein2015hydrogen}%
  \BibitemOpen
  \bibfield  {author} {\bibinfo {author} {\bibfnamefont {J.}~\bibnamefont {Sakstein}},\ }\href@noop {} {\bibfield  {journal} {\bibinfo  {journal} {Physical review letters}\ }\textbf {\bibinfo {volume} {115}},\ \bibinfo {pages} {201101} (\bibinfo {year} {2015}{\natexlab{b}})}\BibitemShut {NoStop}%
\bibitem [{\citenamefont {Olmo}\ \emph {et~al.}(2019)\citenamefont {Olmo}, \citenamefont {Rubiera-Garcia},\ and\ \citenamefont {Wojnar}}]{Olmo:2019qsj}%
  \BibitemOpen
  \bibfield  {author} {\bibinfo {author} {\bibfnamefont {G.~J.}\ \bibnamefont {Olmo}}, \bibinfo {author} {\bibfnamefont {D.}~\bibnamefont {Rubiera-Garcia}}, \ and\ \bibinfo {author} {\bibfnamefont {A.}~\bibnamefont {Wojnar}},\ }\href {\doibase 10.1103/PhysRevD.100.044020} {\bibfield  {journal} {\bibinfo  {journal} {Phys. Rev. D}\ }\textbf {\bibinfo {volume} {100}},\ \bibinfo {pages} {044020} (\bibinfo {year} {2019})},\ \Eprint {http://arxiv.org/abs/1906.04629} {arXiv:1906.04629 [gr-qc]} \BibitemShut {NoStop}%
\bibitem [{\citenamefont {Crisostomi}\ \emph {et~al.}(2019)\citenamefont {Crisostomi}, \citenamefont {Lewandowski},\ and\ \citenamefont {Vernizzi}}]{crisostomi2019vainshtein}%
  \BibitemOpen
  \bibfield  {author} {\bibinfo {author} {\bibfnamefont {M.}~\bibnamefont {Crisostomi}}, \bibinfo {author} {\bibfnamefont {M.}~\bibnamefont {Lewandowski}}, \ and\ \bibinfo {author} {\bibfnamefont {F.}~\bibnamefont {Vernizzi}},\ }\href@noop {} {\bibfield  {journal} {\bibinfo  {journal} {Physical Review D}\ }\textbf {\bibinfo {volume} {100}},\ \bibinfo {pages} {024025} (\bibinfo {year} {2019})}\BibitemShut {NoStop}%
\bibitem [{\citenamefont {Rosyadi}\ \emph {et~al.}(2019)\citenamefont {Rosyadi}, \citenamefont {Sulaksono}, \citenamefont {Kassim},\ and\ \citenamefont {Yusof}}]{rosyadi2019brown}%
  \BibitemOpen
  \bibfield  {author} {\bibinfo {author} {\bibfnamefont {A.}~\bibnamefont {Rosyadi}}, \bibinfo {author} {\bibfnamefont {A.}~\bibnamefont {Sulaksono}}, \bibinfo {author} {\bibfnamefont {H.~A.}\ \bibnamefont {Kassim}}, \ and\ \bibinfo {author} {\bibfnamefont {N.}~\bibnamefont {Yusof}},\ }\href@noop {} {\bibfield  {journal} {\bibinfo  {journal} {The European Physical Journal C}\ }\textbf {\bibinfo {volume} {79}},\ \bibinfo {pages} {1} (\bibinfo {year} {2019})}\BibitemShut {NoStop}%
\bibitem [{\citenamefont {Wojnar}(2021)}]{wojnar2021lithium}%
  \BibitemOpen
  \bibfield  {author} {\bibinfo {author} {\bibfnamefont {A.}~\bibnamefont {Wojnar}},\ }\href@noop {} {\bibfield  {journal} {\bibinfo  {journal} {Physical Review D}\ }\textbf {\bibinfo {volume} {103}},\ \bibinfo {pages} {044037} (\bibinfo {year} {2021})}\BibitemShut {NoStop}%
\bibitem [{\citenamefont {Lecca}(2021)}]{lecca2021effects}%
  \BibitemOpen
  \bibfield  {author} {\bibinfo {author} {\bibfnamefont {P.}~\bibnamefont {Lecca}},\ }in\ \href@noop {} {\emph {\bibinfo {booktitle} {Journal of Physics: Conference Series}}},\ Vol.\ \bibinfo {volume} {2090}\ (\bibinfo {organization} {IOP Publishing},\ \bibinfo {year} {2021})\ p.\ \bibinfo {pages} {012034}\BibitemShut {NoStop}%
\bibitem [{\citenamefont {Hossain}\ and\ \citenamefont {Mandal}(2021{\natexlab{a}})}]{hossain2021equation}%
  \BibitemOpen
  \bibfield  {author} {\bibinfo {author} {\bibfnamefont {G.~M.}\ \bibnamefont {Hossain}}\ and\ \bibinfo {author} {\bibfnamefont {S.}~\bibnamefont {Mandal}},\ }\href@noop {} {\bibfield  {journal} {\bibinfo  {journal} {Journal of Cosmology and Astroparticle Physics}\ }\textbf {\bibinfo {volume} {2021}},\ \bibinfo {pages} {026} (\bibinfo {year} {2021}{\natexlab{a}})}\BibitemShut {NoStop}%
\bibitem [{\citenamefont {Hossain}\ and\ \citenamefont {Mandal}(2021{\natexlab{b}})}]{hossain2021higher}%
  \BibitemOpen
  \bibfield  {author} {\bibinfo {author} {\bibfnamefont {G.~M.}\ \bibnamefont {Hossain}}\ and\ \bibinfo {author} {\bibfnamefont {S.}~\bibnamefont {Mandal}},\ }\href@noop {} {\bibfield  {journal} {\bibinfo  {journal} {Physical Review D}\ }\textbf {\bibinfo {volume} {104}},\ \bibinfo {pages} {123005} (\bibinfo {year} {2021}{\natexlab{b}})}\BibitemShut {NoStop}%
\bibitem [{\citenamefont {Li}\ \emph {et~al.}(2022)\citenamefont {Li}, \citenamefont {Guo}, \citenamefont {Zhao},\ and\ \citenamefont {He}}]{li2022we}%
  \BibitemOpen
  \bibfield  {author} {\bibinfo {author} {\bibfnamefont {J.}~\bibnamefont {Li}}, \bibinfo {author} {\bibfnamefont {T.}~\bibnamefont {Guo}}, \bibinfo {author} {\bibfnamefont {J.}~\bibnamefont {Zhao}}, \ and\ \bibinfo {author} {\bibfnamefont {L.}~\bibnamefont {He}},\ }\href@noop {} {\bibfield  {journal} {\bibinfo  {journal} {arXiv preprint arXiv:2206.02106}\ } (\bibinfo {year} {2022})}\BibitemShut {NoStop}%
\bibitem [{\citenamefont {Chavanis}(2004)}]{chavanis2004statistical}%
  \BibitemOpen
  \bibfield  {author} {\bibinfo {author} {\bibfnamefont {P.-H.}\ \bibnamefont {Chavanis}},\ }\href@noop {} {\bibfield  {journal} {\bibinfo  {journal} {Physical Review E}\ }\textbf {\bibinfo {volume} {69}},\ \bibinfo {pages} {066126} (\bibinfo {year} {2004})}\BibitemShut {NoStop}%
\bibitem [{\citenamefont {Sakstein}\ \emph {et~al.}(2022)\citenamefont {Sakstein}, \citenamefont {Croon},\ and\ \citenamefont {McDermott}}]{sakstein2022axion}%
  \BibitemOpen
  \bibfield  {author} {\bibinfo {author} {\bibfnamefont {J.}~\bibnamefont {Sakstein}}, \bibinfo {author} {\bibfnamefont {D.}~\bibnamefont {Croon}}, \ and\ \bibinfo {author} {\bibfnamefont {S.~D.}\ \bibnamefont {McDermott}},\ }\href@noop {} {\bibfield  {journal} {\bibinfo  {journal} {Physical Review D}\ }\textbf {\bibinfo {volume} {105}},\ \bibinfo {pages} {095038} (\bibinfo {year} {2022})}\BibitemShut {NoStop}%
\bibitem [{\citenamefont {Moussa}(2015)}]{moussa2015effect}%
  \BibitemOpen
  \bibfield  {author} {\bibinfo {author} {\bibfnamefont {M.}~\bibnamefont {Moussa}},\ }\href@noop {} {\bibfield  {journal} {\bibinfo  {journal} {Advances in High Energy Physics}\ }\textbf {\bibinfo {volume} {2015}} (\bibinfo {year} {2015})}\BibitemShut {NoStop}%
\bibitem [{\citenamefont {Rashidi}(2016)}]{rashidi2016generalized}%
  \BibitemOpen
  \bibfield  {author} {\bibinfo {author} {\bibfnamefont {R.}~\bibnamefont {Rashidi}},\ }\href@noop {} {\bibfield  {journal} {\bibinfo  {journal} {Annals of Physics}\ }\textbf {\bibinfo {volume} {374}},\ \bibinfo {pages} {434} (\bibinfo {year} {2016})}\BibitemShut {NoStop}%
\bibitem [{\citenamefont {Belfaqih}\ \emph {et~al.}(2021)\citenamefont {Belfaqih}, \citenamefont {Maulana},\ and\ \citenamefont {Sulaksono}}]{belfaqih2021white}%
  \BibitemOpen
  \bibfield  {author} {\bibinfo {author} {\bibfnamefont {I.~H.}\ \bibnamefont {Belfaqih}}, \bibinfo {author} {\bibfnamefont {H.}~\bibnamefont {Maulana}}, \ and\ \bibinfo {author} {\bibfnamefont {A.}~\bibnamefont {Sulaksono}},\ }\href@noop {} {\bibfield  {journal} {\bibinfo  {journal} {International Journal of Modern Physics D}\ }\textbf {\bibinfo {volume} {30}},\ \bibinfo {pages} {2150064} (\bibinfo {year} {2021})}\BibitemShut {NoStop}%
\bibitem [{\citenamefont {Mathew}\ and\ \citenamefont {Nandy}(2021)}]{mathew2021existence}%
  \BibitemOpen
  \bibfield  {author} {\bibinfo {author} {\bibfnamefont {A.}~\bibnamefont {Mathew}}\ and\ \bibinfo {author} {\bibfnamefont {M.~K.}\ \bibnamefont {Nandy}},\ }\href@noop {} {\bibfield  {journal} {\bibinfo  {journal} {Royal Society open science}\ }\textbf {\bibinfo {volume} {8}},\ \bibinfo {pages} {210301} (\bibinfo {year} {2021})}\BibitemShut {NoStop}%
\bibitem [{\citenamefont {Hamil}\ and\ \citenamefont {L{\"u}tf{\"u}o{\u{g}}lu}(2021)}]{hamil2021new}%
  \BibitemOpen
  \bibfield  {author} {\bibinfo {author} {\bibfnamefont {B.}~\bibnamefont {Hamil}}\ and\ \bibinfo {author} {\bibfnamefont {B.}~\bibnamefont {L{\"u}tf{\"u}o{\u{g}}lu}},\ }\href@noop {} {\bibfield  {journal} {\bibinfo  {journal} {International Journal of Theoretical Physics}\ }\textbf {\bibinfo {volume} {60}},\ \bibinfo {pages} {2790} (\bibinfo {year} {2021})}\BibitemShut {NoStop}%
\bibitem [{\citenamefont {Gregoris}\ and\ \citenamefont {Ong}(2022)}]{gregoris2022chadrasekhar}%
  \BibitemOpen
  \bibfield  {author} {\bibinfo {author} {\bibfnamefont {D.}~\bibnamefont {Gregoris}}\ and\ \bibinfo {author} {\bibfnamefont {Y.~C.}\ \bibnamefont {Ong}},\ }\href@noop {} {\bibfield  {journal} {\bibinfo  {journal} {arXiv preprint arXiv:2202.13904}\ } (\bibinfo {year} {2022})}\BibitemShut {NoStop}%
\bibitem [{\citenamefont {Pacho\l{}}\ and\ \citenamefont {Wojnar}(2023{\natexlab{a}})}]{Pachol:2023bkv}%
  \BibitemOpen
  \bibfield  {author} {\bibinfo {author} {\bibfnamefont {A.}~\bibnamefont {Pacho\l{}}}\ and\ \bibinfo {author} {\bibfnamefont {A.}~\bibnamefont {Wojnar}},\ }\href@noop {} {\  (\bibinfo {year} {2023}{\natexlab{a}})},\ \Eprint {http://arxiv.org/abs/2307.03520} {arXiv:2307.03520 [gr-qc]} \BibitemShut {NoStop}%
\bibitem [{\citenamefont {Pacho\l{}}\ and\ \citenamefont {Wojnar}(2023{\natexlab{b}})}]{Pachol:2023tqa}%
  \BibitemOpen
  \bibfield  {author} {\bibinfo {author} {\bibfnamefont {A.}~\bibnamefont {Pacho\l{}}}\ and\ \bibinfo {author} {\bibfnamefont {A.}~\bibnamefont {Wojnar}},\ }\href {\doibase 10.1088/1361-6382/acf435} {\bibfield  {journal} {\bibinfo  {journal} {Class. Quant. Grav.}\ }\textbf {\bibinfo {volume} {40}},\ \bibinfo {pages} {195021} (\bibinfo {year} {2023}{\natexlab{b}})},\ \Eprint {http://arxiv.org/abs/2304.08215} {arXiv:2304.08215 [gr-qc]} \BibitemShut {NoStop}%
\bibitem [{\citenamefont {Kozak}\ \emph {et~al.}(2023)\citenamefont {Kozak}, \citenamefont {Pacho\l{}},\ and\ \citenamefont {Wojnar}}]{Kozak:2023vlj}%
  \BibitemOpen
  \bibfield  {author} {\bibinfo {author} {\bibfnamefont {A.}~\bibnamefont {Kozak}}, \bibinfo {author} {\bibfnamefont {A.}~\bibnamefont {Pacho\l{}}}, \ and\ \bibinfo {author} {\bibfnamefont {A.}~\bibnamefont {Wojnar}},\ }\href@noop {} {\  (\bibinfo {year} {2023})},\ \Eprint {http://arxiv.org/abs/2310.00913} {arXiv:2310.00913 [gr-qc]} \BibitemShut {NoStop}%
\bibitem [{\citenamefont {Kempf}\ \emph {et~al.}(1995)\citenamefont {Kempf}, \citenamefont {Mangano},\ and\ \citenamefont {Mann}}]{kempf1995hilbert}%
  \BibitemOpen
  \bibfield  {author} {\bibinfo {author} {\bibfnamefont {A.}~\bibnamefont {Kempf}}, \bibinfo {author} {\bibfnamefont {G.}~\bibnamefont {Mangano}}, \ and\ \bibinfo {author} {\bibfnamefont {R.~B.}\ \bibnamefont {Mann}},\ }\href@noop {} {\bibfield  {journal} {\bibinfo  {journal} {Physical Review D}\ }\textbf {\bibinfo {volume} {52}},\ \bibinfo {pages} {1108} (\bibinfo {year} {1995})}\BibitemShut {NoStop}%
\bibitem [{\citenamefont {Maggiore}(1993)}]{maggiore1993generalized}%
  \BibitemOpen
  \bibfield  {author} {\bibinfo {author} {\bibfnamefont {M.}~\bibnamefont {Maggiore}},\ }\href@noop {} {\bibfield  {journal} {\bibinfo  {journal} {Physics Letters B}\ }\textbf {\bibinfo {volume} {304}},\ \bibinfo {pages} {65} (\bibinfo {year} {1993})}\BibitemShut {NoStop}%
\bibitem [{\citenamefont {Maggiore}(1994)}]{maggiore1994quantum}%
  \BibitemOpen
  \bibfield  {author} {\bibinfo {author} {\bibfnamefont {M.}~\bibnamefont {Maggiore}},\ }\href@noop {} {\bibfield  {journal} {\bibinfo  {journal} {Physical Review D}\ }\textbf {\bibinfo {volume} {49}},\ \bibinfo {pages} {5182} (\bibinfo {year} {1994})}\BibitemShut {NoStop}%
\bibitem [{\citenamefont {Chang}\ \emph {et~al.}(2002{\natexlab{a}})\citenamefont {Chang}, \citenamefont {Minic}, \citenamefont {Okamura},\ and\ \citenamefont {Takeuchi}}]{chang2002exact}%
  \BibitemOpen
  \bibfield  {author} {\bibinfo {author} {\bibfnamefont {L.~N.}\ \bibnamefont {Chang}}, \bibinfo {author} {\bibfnamefont {D.}~\bibnamefont {Minic}}, \bibinfo {author} {\bibfnamefont {N.}~\bibnamefont {Okamura}}, \ and\ \bibinfo {author} {\bibfnamefont {T.}~\bibnamefont {Takeuchi}},\ }\href@noop {} {\bibfield  {journal} {\bibinfo  {journal} {Physical Review D}\ }\textbf {\bibinfo {volume} {65}},\ \bibinfo {pages} {125027} (\bibinfo {year} {2002}{\natexlab{a}})}\BibitemShut {NoStop}%
\bibitem [{\citenamefont {Chang}\ \emph {et~al.}(2002{\natexlab{b}})\citenamefont {Chang}, \citenamefont {Minic}, \citenamefont {Okamura},\ and\ \citenamefont {Takeuchi}}]{chang2002effect}%
  \BibitemOpen
  \bibfield  {author} {\bibinfo {author} {\bibfnamefont {L.~N.}\ \bibnamefont {Chang}}, \bibinfo {author} {\bibfnamefont {D.}~\bibnamefont {Minic}}, \bibinfo {author} {\bibfnamefont {N.}~\bibnamefont {Okamura}}, \ and\ \bibinfo {author} {\bibfnamefont {T.}~\bibnamefont {Takeuchi}},\ }\href@noop {} {\bibfield  {journal} {\bibinfo  {journal} {Physical Review D}\ }\textbf {\bibinfo {volume} {65}},\ \bibinfo {pages} {125028} (\bibinfo {year} {2002}{\natexlab{b}})}\BibitemShut {NoStop}%
\bibitem [{\citenamefont {Bishop}\ \emph {et~al.}(2020)\citenamefont {Bishop}, \citenamefont {Lee},\ and\ \citenamefont {Singleton}}]{bishop2020modified}%
  \BibitemOpen
  \bibfield  {author} {\bibinfo {author} {\bibfnamefont {M.}~\bibnamefont {Bishop}}, \bibinfo {author} {\bibfnamefont {J.}~\bibnamefont {Lee}}, \ and\ \bibinfo {author} {\bibfnamefont {D.}~\bibnamefont {Singleton}},\ }\href@noop {} {\bibfield  {journal} {\bibinfo  {journal} {Physics Letters B}\ }\textbf {\bibinfo {volume} {802}},\ \bibinfo {pages} {135209} (\bibinfo {year} {2020})}\BibitemShut {NoStop}%
\bibitem [{\citenamefont {Bishop}\ \emph {et~al.}(2022)\citenamefont {Bishop}, \citenamefont {Contreras},\ and\ \citenamefont {Singleton}}]{bishop2022subtle}%
  \BibitemOpen
  \bibfield  {author} {\bibinfo {author} {\bibfnamefont {M.}~\bibnamefont {Bishop}}, \bibinfo {author} {\bibfnamefont {J.}~\bibnamefont {Contreras}}, \ and\ \bibinfo {author} {\bibfnamefont {D.}~\bibnamefont {Singleton}},\ }\href@noop {} {\bibfield  {journal} {\bibinfo  {journal} {universe}\ }\textbf {\bibinfo {volume} {8}},\ \bibinfo {pages} {192} (\bibinfo {year} {2022})}\BibitemShut {NoStop}%
\bibitem [{\citenamefont {Segreto}\ and\ \citenamefont {Montani}(2023)}]{segreto2023extended}%
  \BibitemOpen
  \bibfield  {author} {\bibinfo {author} {\bibfnamefont {S.}~\bibnamefont {Segreto}}\ and\ \bibinfo {author} {\bibfnamefont {G.}~\bibnamefont {Montani}},\ }\href@noop {} {\bibfield  {journal} {\bibinfo  {journal} {The European Physical Journal C}\ }\textbf {\bibinfo {volume} {83}},\ \bibinfo {pages} {385} (\bibinfo {year} {2023})}\BibitemShut {NoStop}%
\bibitem [{\citenamefont {Wojnar}(2023{\natexlab{b}})}]{Wojnar:2023bvv}%
  \BibitemOpen
  \bibfield  {author} {\bibinfo {author} {\bibfnamefont {A.}~\bibnamefont {Wojnar}},\ }\href@noop {} {\  (\bibinfo {year} {2023}{\natexlab{b}})},\ \Eprint {http://arxiv.org/abs/2311.14066} {arXiv:2311.14066 [gr-qc]} \BibitemShut {NoStop}%
\bibitem [{\citenamefont {Alfonso}\ \emph {et~al.}(2017)\citenamefont {Alfonso}, \citenamefont {Bejarano}, \citenamefont {Jimenez}, \citenamefont {Olmo},\ and\ \citenamefont {Orazi}}]{alfonso2017trivial}%
  \BibitemOpen
  \bibfield  {author} {\bibinfo {author} {\bibfnamefont {V.~I.}\ \bibnamefont {Alfonso}}, \bibinfo {author} {\bibfnamefont {C.}~\bibnamefont {Bejarano}}, \bibinfo {author} {\bibfnamefont {J.~B.}\ \bibnamefont {Jimenez}}, \bibinfo {author} {\bibfnamefont {G.~J.}\ \bibnamefont {Olmo}}, \ and\ \bibinfo {author} {\bibfnamefont {E.}~\bibnamefont {Orazi}},\ }\href@noop {} {\bibfield  {journal} {\bibinfo  {journal} {Classical and Quantum Gravity}\ }\textbf {\bibinfo {volume} {34}},\ \bibinfo {pages} {235003} (\bibinfo {year} {2017})}\BibitemShut {NoStop}%
\bibitem [{\citenamefont {Borowiec}\ \emph {et~al.}(1998)\citenamefont {Borowiec}, \citenamefont {Ferraris}, \citenamefont {Francaviglia},\ and\ \citenamefont {Volovich}}]{Borowiec:1996kg}%
  \BibitemOpen
  \bibfield  {author} {\bibinfo {author} {\bibfnamefont {A.}~\bibnamefont {Borowiec}}, \bibinfo {author} {\bibfnamefont {M.}~\bibnamefont {Ferraris}}, \bibinfo {author} {\bibfnamefont {M.}~\bibnamefont {Francaviglia}}, \ and\ \bibinfo {author} {\bibfnamefont {I.}~\bibnamefont {Volovich}},\ }\href {\doibase 10.1088/0264-9381/15/1/005} {\bibfield  {journal} {\bibinfo  {journal} {Class. Quant. Grav.}\ }\textbf {\bibinfo {volume} {15}},\ \bibinfo {pages} {43} (\bibinfo {year} {1998})},\ \Eprint {http://arxiv.org/abs/gr-qc/9611067} {arXiv:gr-qc/9611067} \BibitemShut {NoStop}%
\bibitem [{\citenamefont {Allemandi}\ \emph {et~al.}(2004)\citenamefont {Allemandi}, \citenamefont {Borowiec},\ and\ \citenamefont {Francaviglia}}]{Allemandi:2004wn}%
  \BibitemOpen
  \bibfield  {author} {\bibinfo {author} {\bibfnamefont {G.}~\bibnamefont {Allemandi}}, \bibinfo {author} {\bibfnamefont {A.}~\bibnamefont {Borowiec}}, \ and\ \bibinfo {author} {\bibfnamefont {M.}~\bibnamefont {Francaviglia}},\ }\href {\doibase 10.1103/PhysRevD.70.103503} {\bibfield  {journal} {\bibinfo  {journal} {Phys. Rev. D}\ }\textbf {\bibinfo {volume} {70}},\ \bibinfo {pages} {103503} (\bibinfo {year} {2004})},\ \Eprint {http://arxiv.org/abs/hep-th/0407090} {arXiv:hep-th/0407090} \BibitemShut {NoStop}%
\bibitem [{\citenamefont {Beltr{\'a}n~Jim{\'e}nez}\ and\ \citenamefont {Delhom}(2019)}]{beltran2019ghosts}%
  \BibitemOpen
  \bibfield  {author} {\bibinfo {author} {\bibfnamefont {J.}~\bibnamefont {Beltr{\'a}n~Jim{\'e}nez}}\ and\ \bibinfo {author} {\bibfnamefont {A.}~\bibnamefont {Delhom}},\ }\href@noop {} {\bibfield  {journal} {\bibinfo  {journal} {The European Physical Journal C}\ }\textbf {\bibinfo {volume} {79}},\ \bibinfo {pages} {1} (\bibinfo {year} {2019})}\BibitemShut {NoStop}%
\bibitem [{\citenamefont {Jim{\'e}nez}\ and\ \citenamefont {Delhom}(2020)}]{jimenez2020instabilities}%
  \BibitemOpen
  \bibfield  {author} {\bibinfo {author} {\bibfnamefont {J.~B.}\ \bibnamefont {Jim{\'e}nez}}\ and\ \bibinfo {author} {\bibfnamefont {A.}~\bibnamefont {Delhom}},\ }\href@noop {} {\bibfield  {journal} {\bibinfo  {journal} {The European Physical Journal C}\ }\textbf {\bibinfo {volume} {80}},\ \bibinfo {pages} {585} (\bibinfo {year} {2020})}\BibitemShut {NoStop}%
\bibitem [{\citenamefont {Vollick}(2004)}]{vollick2004palatini}%
  \BibitemOpen
  \bibfield  {author} {\bibinfo {author} {\bibfnamefont {D.~N.}\ \bibnamefont {Vollick}},\ }\href@noop {} {\bibfield  {journal} {\bibinfo  {journal} {Physical Review D}\ }\textbf {\bibinfo {volume} {69}},\ \bibinfo {pages} {064030} (\bibinfo {year} {2004})}\BibitemShut {NoStop}%
\bibitem [{\citenamefont {Jim{\'e}nez}\ \emph {et~al.}(2018)\citenamefont {Jim{\'e}nez}, \citenamefont {Heisenberg}, \citenamefont {Olmo},\ and\ \citenamefont {Rubiera-Garcia}}]{jimenez2018born}%
  \BibitemOpen
  \bibfield  {author} {\bibinfo {author} {\bibfnamefont {J.~B.}\ \bibnamefont {Jim{\'e}nez}}, \bibinfo {author} {\bibfnamefont {L.}~\bibnamefont {Heisenberg}}, \bibinfo {author} {\bibfnamefont {G.~J.}\ \bibnamefont {Olmo}}, \ and\ \bibinfo {author} {\bibfnamefont {D.}~\bibnamefont {Rubiera-Garcia}},\ }\href@noop {} {\bibfield  {journal} {\bibinfo  {journal} {Physics Reports}\ }\textbf {\bibinfo {volume} {727}},\ \bibinfo {pages} {1} (\bibinfo {year} {2018})}\BibitemShut {NoStop}%
\bibitem [{\citenamefont {Toniato}\ \emph {et~al.}(2020)\citenamefont {Toniato}, \citenamefont {Rodrigues},\ and\ \citenamefont {Wojnar}}]{Toniato:2019rrd}%
  \BibitemOpen
  \bibfield  {author} {\bibinfo {author} {\bibfnamefont {J.~D.}\ \bibnamefont {Toniato}}, \bibinfo {author} {\bibfnamefont {D.~C.}\ \bibnamefont {Rodrigues}}, \ and\ \bibinfo {author} {\bibfnamefont {A.}~\bibnamefont {Wojnar}},\ }\href {\doibase 10.1103/PhysRevD.101.064050} {\bibfield  {journal} {\bibinfo  {journal} {Phys. Rev. D}\ }\textbf {\bibinfo {volume} {101}},\ \bibinfo {pages} {064050} (\bibinfo {year} {2020})},\ \Eprint {http://arxiv.org/abs/1912.12234} {arXiv:1912.12234 [gr-qc]} \BibitemShut {NoStop}%
\bibitem [{\citenamefont {Banados}\ and\ \citenamefont {Ferreira}(2010)}]{banados2010eddington}%
  \BibitemOpen
  \bibfield  {author} {\bibinfo {author} {\bibfnamefont {M.}~\bibnamefont {Banados}}\ and\ \bibinfo {author} {\bibfnamefont {P.~G.}\ \bibnamefont {Ferreira}},\ }\href@noop {} {\bibfield  {journal} {\bibinfo  {journal} {Physical review letters}\ }\textbf {\bibinfo {volume} {105}},\ \bibinfo {pages} {011101} (\bibinfo {year} {2010})}\BibitemShut {NoStop}%
\bibitem [{\citenamefont {Pani}\ \emph {et~al.}(2011)\citenamefont {Pani}, \citenamefont {Cardoso},\ and\ \citenamefont {Delsate}}]{pani2011compact}%
  \BibitemOpen
  \bibfield  {author} {\bibinfo {author} {\bibfnamefont {P.}~\bibnamefont {Pani}}, \bibinfo {author} {\bibfnamefont {V.}~\bibnamefont {Cardoso}}, \ and\ \bibinfo {author} {\bibfnamefont {T.}~\bibnamefont {Delsate}},\ }\href@noop {} {\bibfield  {journal} {\bibinfo  {journal} {Physical Review Letters}\ }\textbf {\bibinfo {volume} {107}},\ \bibinfo {pages} {031101} (\bibinfo {year} {2011})}\BibitemShut {NoStop}%
\bibitem [{\citenamefont {Pani}\ and\ \citenamefont {Sotiriou}(2012)}]{pani2012surface}%
  \BibitemOpen
  \bibfield  {author} {\bibinfo {author} {\bibfnamefont {P.}~\bibnamefont {Pani}}\ and\ \bibinfo {author} {\bibfnamefont {T.~P.}\ \bibnamefont {Sotiriou}},\ }\href@noop {} {\bibfield  {journal} {\bibinfo  {journal} {Physical review letters}\ }\textbf {\bibinfo {volume} {109}},\ \bibinfo {pages} {251102} (\bibinfo {year} {2012})}\BibitemShut {NoStop}%
\bibitem [{\citenamefont {Cortes}\ and\ \citenamefont {Gamboa}(2020)}]{cortes2020deformed}%
  \BibitemOpen
  \bibfield  {author} {\bibinfo {author} {\bibfnamefont {J.~L.}\ \bibnamefont {Cortes}}\ and\ \bibinfo {author} {\bibfnamefont {J.}~\bibnamefont {Gamboa}},\ }\href@noop {} {\bibfield  {journal} {\bibinfo  {journal} {Physical Review D}\ }\textbf {\bibinfo {volume} {102}},\ \bibinfo {pages} {036015} (\bibinfo {year} {2020})}\BibitemShut {NoStop}%
\bibitem [{\citenamefont {Ali}\ \emph {et~al.}(2009)\citenamefont {Ali}, \citenamefont {Das},\ and\ \citenamefont {Vagenas}}]{ali2009discreteness}%
  \BibitemOpen
  \bibfield  {author} {\bibinfo {author} {\bibfnamefont {A.~F.}\ \bibnamefont {Ali}}, \bibinfo {author} {\bibfnamefont {S.}~\bibnamefont {Das}}, \ and\ \bibinfo {author} {\bibfnamefont {E.~C.}\ \bibnamefont {Vagenas}},\ }\href@noop {} {\bibfield  {journal} {\bibinfo  {journal} {Physics Letters B}\ }\textbf {\bibinfo {volume} {678}},\ \bibinfo {pages} {497} (\bibinfo {year} {2009})}\BibitemShut {NoStop}%
\bibitem [{\citenamefont {Ali}(2011)}]{ali2011minimal}%
  \BibitemOpen
  \bibfield  {author} {\bibinfo {author} {\bibfnamefont {A.~F.}\ \bibnamefont {Ali}},\ }\href@noop {} {\bibfield  {journal} {\bibinfo  {journal} {Classical and Quantum Gravity}\ }\textbf {\bibinfo {volume} {28}},\ \bibinfo {pages} {065013} (\bibinfo {year} {2011})}\BibitemShut {NoStop}%
\bibitem [{\citenamefont {Abac}\ \emph {et~al.}(2021)\citenamefont {Abac}, \citenamefont {Esguerra},\ and\ \citenamefont {Otadoy}}]{abac2021modified}%
  \BibitemOpen
  \bibfield  {author} {\bibinfo {author} {\bibfnamefont {A.~G.}\ \bibnamefont {Abac}}, \bibinfo {author} {\bibfnamefont {J.~P.~H.}\ \bibnamefont {Esguerra}}, \ and\ \bibinfo {author} {\bibfnamefont {R.~E.~S.}\ \bibnamefont {Otadoy}},\ }\href@noop {} {\bibfield  {journal} {\bibinfo  {journal} {International Journal of Modern Physics D}\ }\textbf {\bibinfo {volume} {30}},\ \bibinfo {pages} {2150005} (\bibinfo {year} {2021})}\BibitemShut {NoStop}%
\bibitem [{\citenamefont {Vagenas}\ \emph {et~al.}(2019{\natexlab{a}})\citenamefont {Vagenas}, \citenamefont {Farag~Ali},\ and\ \citenamefont {Alshal}}]{vagenas2019gup}%
  \BibitemOpen
  \bibfield  {author} {\bibinfo {author} {\bibfnamefont {E.~C.}\ \bibnamefont {Vagenas}}, \bibinfo {author} {\bibfnamefont {A.}~\bibnamefont {Farag~Ali}}, \ and\ \bibinfo {author} {\bibfnamefont {H.}~\bibnamefont {Alshal}},\ }\href@noop {} {\bibfield  {journal} {\bibinfo  {journal} {The European Physical Journal C}\ }\textbf {\bibinfo {volume} {79}},\ \bibinfo {pages} {1} (\bibinfo {year} {2019}{\natexlab{a}})}\BibitemShut {NoStop}%
\bibitem [{\citenamefont {Tawfik}\ and\ \citenamefont {Diab}(2014)}]{tawfik2014generalized}%
  \BibitemOpen
  \bibfield  {author} {\bibinfo {author} {\bibfnamefont {A.}~\bibnamefont {Tawfik}}\ and\ \bibinfo {author} {\bibfnamefont {A.}~\bibnamefont {Diab}},\ }\href@noop {} {\bibfield  {journal} {\bibinfo  {journal} {International Journal of Modern Physics D}\ }\textbf {\bibinfo {volume} {23}},\ \bibinfo {pages} {1430025} (\bibinfo {year} {2014})}\BibitemShut {NoStop}%
\bibitem [{\citenamefont {Vagenas}\ \emph {et~al.}(2019{\natexlab{b}})\citenamefont {Vagenas}, \citenamefont {Ali}, \citenamefont {Hemeda},\ and\ \citenamefont {Alshal}}]{vagenas2019linear}%
  \BibitemOpen
  \bibfield  {author} {\bibinfo {author} {\bibfnamefont {E.~C.}\ \bibnamefont {Vagenas}}, \bibinfo {author} {\bibfnamefont {A.~F.}\ \bibnamefont {Ali}}, \bibinfo {author} {\bibfnamefont {M.}~\bibnamefont {Hemeda}}, \ and\ \bibinfo {author} {\bibfnamefont {H.}~\bibnamefont {Alshal}},\ }\href@noop {} {\bibfield  {journal} {\bibinfo  {journal} {The European Physical Journal C}\ }\textbf {\bibinfo {volume} {79}},\ \bibinfo {pages} {1} (\bibinfo {year} {2019}{\natexlab{b}})}\BibitemShut {NoStop}%
\bibitem [{\citenamefont {Huang}(2009)}]{huang2009introduction}%
  \BibitemOpen
  \bibfield  {author} {\bibinfo {author} {\bibfnamefont {K.}~\bibnamefont {Huang}},\ }\href@noop {} {\emph {\bibinfo {title} {Introduction to statistical physics}}}\ (\bibinfo  {publisher} {CRC press},\ \bibinfo {year} {2009})\BibitemShut {NoStop}%
\bibitem [{\citenamefont {Kozak}\ and\ \citenamefont {Wojnar}(2023{\natexlab{a}})}]{Kozak:2023axy}%
  \BibitemOpen
  \bibfield  {author} {\bibinfo {author} {\bibfnamefont {A.}~\bibnamefont {Kozak}}\ and\ \bibinfo {author} {\bibfnamefont {A.}~\bibnamefont {Wojnar}},\ }\href {\doibase 10.1103/PhysRevD.108.044055} {\bibfield  {journal} {\bibinfo  {journal} {Phys. Rev. D}\ }\textbf {\bibinfo {volume} {108}},\ \bibinfo {pages} {044055} (\bibinfo {year} {2023}{\natexlab{a}})},\ \Eprint {http://arxiv.org/abs/2303.17213} {arXiv:2303.17213 [gr-qc]} \BibitemShut {NoStop}%
\bibitem [{\citenamefont {Kozak}\ and\ \citenamefont {Wojnar}(2023{\natexlab{b}})}]{Kozak:2023ruu}%
  \BibitemOpen
  \bibfield  {author} {\bibinfo {author} {\bibfnamefont {A.}~\bibnamefont {Kozak}}\ and\ \bibinfo {author} {\bibfnamefont {A.}~\bibnamefont {Wojnar}},\ }\href@noop {} {\  (\bibinfo {year} {2023}{\natexlab{b}})},\ \Eprint {http://arxiv.org/abs/2308.01784} {arXiv:2308.01784 [gr-qc]} \BibitemShut {NoStop}%
\bibitem [{\citenamefont {Landau}(2018)}]{landau2018theory}%
  \BibitemOpen
  \bibfield  {author} {\bibinfo {author} {\bibfnamefont {L.}~\bibnamefont {Landau}},\ }in\ \href@noop {} {\emph {\bibinfo {booktitle} {An Introduction to the Theory of Superfluidity}}}\ (\bibinfo  {publisher} {CRC Press},\ \bibinfo {year} {2018})\ pp.\ \bibinfo {pages} {185--204}\BibitemShut {NoStop}%
\bibitem [{\citenamefont {Tisza}(1947)}]{tisza1947theory}%
  \BibitemOpen
  \bibfield  {author} {\bibinfo {author} {\bibfnamefont {L.}~\bibnamefont {Tisza}},\ }\href@noop {} {\bibfield  {journal} {\bibinfo  {journal} {Physical Review}\ }\textbf {\bibinfo {volume} {72}},\ \bibinfo {pages} {838} (\bibinfo {year} {1947})}\BibitemShut {NoStop}%
\bibitem [{\citenamefont {Cohen}\ and\ \citenamefont {Feynman}(1957)}]{cohen1957theory}%
  \BibitemOpen
  \bibfield  {author} {\bibinfo {author} {\bibfnamefont {M.}~\bibnamefont {Cohen}}\ and\ \bibinfo {author} {\bibfnamefont {R.~P.}\ \bibnamefont {Feynman}},\ }\href@noop {} {\bibfield  {journal} {\bibinfo  {journal} {Physical Review}\ }\textbf {\bibinfo {volume} {107}},\ \bibinfo {pages} {13} (\bibinfo {year} {1957})}\BibitemShut {NoStop}%
\bibitem [{\citenamefont {Yarnell}\ \emph {et~al.}(1959)\citenamefont {Yarnell}, \citenamefont {Arnold}, \citenamefont {Bendt},\ and\ \citenamefont {Kerr}}]{yarnell1959excitations}%
  \BibitemOpen
  \bibfield  {author} {\bibinfo {author} {\bibfnamefont {J.}~\bibnamefont {Yarnell}}, \bibinfo {author} {\bibfnamefont {G.}~\bibnamefont {Arnold}}, \bibinfo {author} {\bibfnamefont {P.}~\bibnamefont {Bendt}}, \ and\ \bibinfo {author} {\bibfnamefont {E.}~\bibnamefont {Kerr}},\ }\href@noop {} {\bibfield  {journal} {\bibinfo  {journal} {Physical Review}\ }\textbf {\bibinfo {volume} {113}},\ \bibinfo {pages} {1379} (\bibinfo {year} {1959})}\BibitemShut {NoStop}%
\bibitem [{\citenamefont {Kramers}(1957)}]{kramers1957chapter}%
  \BibitemOpen
  \bibfield  {author} {\bibinfo {author} {\bibfnamefont {H.}~\bibnamefont {Kramers}},\ }in\ \href@noop {} {\emph {\bibinfo {booktitle} {Progress in Low Temperature Physics}}},\ Vol.~\bibinfo {volume} {2}\ (\bibinfo  {publisher} {Elsevier},\ \bibinfo {year} {1957})\ pp.\ \bibinfo {pages} {59--82}\BibitemShut {NoStop}%
\end{thebibliography}%

\end{document}